\documentclass[12pt]{article}
\usepackage{color,amssymb,epsfig,verbatim}
\usepackage{graphicx} 
\usepackage{bm}
\usepackage{multirow}
\usepackage{amsmath}
\usepackage{lscape}
\usepackage[table,xcdraw]{xcolor}
\usepackage{natbib}
\setcitestyle{aysep={}} 
\usepackage{hyperref}
\usepackage{accents}
\hypersetup{colorlinks,linkcolor={black},citecolor={black},urlcolor={red}}

\def\boxit#1{\vbox{\hrule\hbox{\vrule\kern6pt
          \vbox{\kern6pt#1\kern6pt}\kern6pt\vrule}\hrule}}

\def\bse{\begin{eqnarray*}}
\def\ese{\end{eqnarray*}}
\def\be{\begin{eqnarray}}
\def\ee{\end{eqnarray}}
\def\bq{\begin{equation}}
\def\eq{\end{equation}}
\def\bse{\begin{eqnarray*}}
\def\ese{\end{eqnarray*}}

\definecolor{emerald}{RGB}{6, 91, 82}

\usepackage{enumitem}

\usepackage{scalerel}

\usepackage{rotating} 
\usepackage[para]{threeparttable}
\usepackage{caption}

\usepackage{ dsfont }

\usepackage{romannum}

\newcommand{\norm}[1]{\left\lVert#1\right\rVert}

\begin{document}
\thispagestyle{empty}

\hfill\today \\ \\

\baselineskip=28pt
\begin{center}
{\LARGE{\bf Text Mining and Sentiment Analysis of COVID-19 Tweets}}
\end{center}
\baselineskip=14pt
\vskip 2mm
\begin{center}
Qihuang Zhang\footnote{\baselineskip=10pt  Department of Statistical and Actuarial Sciences, University of Western Ontario, London, Ontario, Canada}, Grace Y. Yi$^{1}$\footnote{\baselineskip=10pt Department of Computer Science, University of Western Ontario, London, Ontario, Canada}\footnote{\baselineskip=10pt Corresponding author. Email: gyi5@uwo.ca}, Li-Pang Chen$^{1}$ and Wenqing He$^{1}$

\end{center}
\bigskip

\vspace{8mm}

\begin{center}
{\Large{\bf Abstract}}
\end{center}
\baselineskip=17pt
{ The human severe acute respiratory syndrome coronavirus 2 (SARS-Cov-2), causing the COVID-19 disease, has continued to spread all over the world. It menacingly affects not only public health and global economics but also mental health and mood. While the impact of the COVID-19 pandemic has been widely studied, relatively fewer discussions about the sentimental reaction of the population have been available. In this article, we scrape \mbox{COVID-19} related tweets on the microblogging platform, Twitter, and examine the tweets from Feb~24, 2020 to Oct~14, 2020 in four Canadian cities (Toronto, Montreal, Vancouver, and Calgary) and four U.S. cities (New York, Los Angeles, Chicago, and Seattle). Applying the Vader and NRC approaches, we evaluate the sentiment intensity scores and visualize the information over different periods of the pandemic. Sentiment scores for the tweets concerning three anti-epidemic measures, masks, vaccine, and lockdown, are computed for comparisons. The results of four Canadian cities are compared with four cities in the United States. We study the causal relationships between the infected cases, the tweet activities, and the sentiment scores of COVID-19 related tweets, by integrating the echo state network method with convergent cross-mapping. Our analysis shows that public sentiments regarding COVID-19 vary in different time periods and locations. In general, people have a positive mood about COVID-19 and masks, but negative in the topics of vaccine and lockdown. The causal inference shows that the sentiment influences people's activities on Twitter, which is also correlated to the daily number of infections. 

}

\vspace{8mm}

\par\vfill\noindent
\underline{\bf Keywords}:  COVID-19, Causal Inference, Echo State Network, Emotion,  Sentiment Analysis, Text Mining.

\par\medskip\noindent
\underline{\bf Short title}: Sentiment Analysis  of  COVID-19 Tweets {\color{white} SentiTwi}

\clearpage\pagebreak\newpage
\pagenumbering{arabic}

\newlength{\gnat}
\setlength{\gnat}{22pt}
\baselineskip=\gnat

\clearpage



\section{Introduction} 

The COVID-19 disease, caused by the human severe acute respiratory syndrome coronavirus~2 (SARS-Cov-2), was declared to be a pandemic by the World Health Organization (WHO) in March of 2020. This disease has caused over sixty-five million infections and a half million deaths all over the world as of December 3, 2020. While extensive studies have been conducted to examine various types of influence of COVID-19 on public health, including the studies concerning the infected cases number and the fatality, investigations of the impact of the pandemic on people's emotion are relatively limited.

\begin{figure}[h]
 \centering
 \makebox[10pt]{%
    \includegraphics[width=0.8\paperwidth]{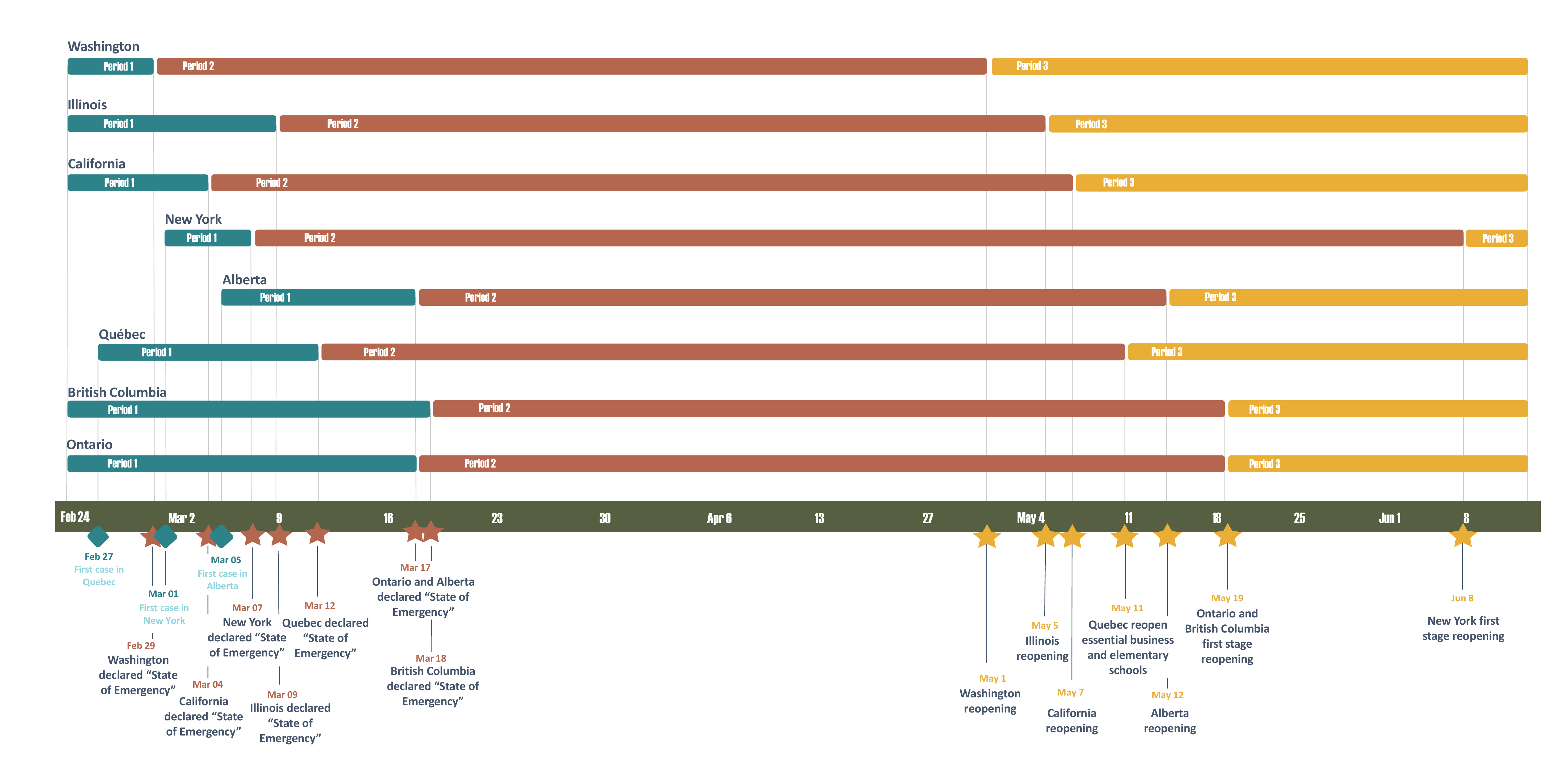}}
 \caption{The timeline of the measures, indicated by the three periods, which are taken by the four provincial governments in Canada: Alberta, Quebec, British Columbia and Ontario and four state governments in U.S.: Washington, Illinois, California, and New York.}\label{fig:timeline}
\end{figure}

The first case of COVID-19 in Canada was reported in Toronto on January 25, 2020. To prevent the spread of COVID-19, four most populated provinces in Canada have consecutively announced the ``state of emergency" (displayed in Figure~\ref{fig:timeline}), taking the measures of shutting down the public business, banning social gathering, encouraging social distancing, and requiring the masks wearing in the public area, etc. As the situation of disease spreading ameliorated during July and August, the ``state of emergency" was relaxed to various extent in different provinces. With the recent roaring number of newly infected cases, the ``state of emergence" has been restored again in all the four provinces.

While social distancing has been advocated and many people are taking the practice of working at home, online communication tools such as social media have become active in communication. The COVID-19 pandemic has become one of the most discussed topics on the internet since it was first reported in January 2020 \citep{bhat2020sentiment}. As the opinions and feelings are freely and openly shared on the internet, it is interesting to conduct text mining of the public information on the social media to extract useful messages. Text mining is a commonly used technique to explore corpus, and common strategies of analyzing sentiments can be found in \citet[][Ch.4]{kwartler2017text}. For an application of text mining to Twitter data, \citet{kumar2014twitter} provided a comprehensive discussion together with detailed case studies. \citet{aflakparast2020analysis} used the Bayesian fused graphical lasso to convert textual Twitter data to understandable networks of terms that can signify important events. 

For COVID-19 data, \citet{tworowski2020covid19} studied drug repository and applied text mining methods to putative COVID-19 therapeutics. \citet{khanday2020machine} employed text mining methods to do preprocessing and extract relevant features, and then classified textual clinical reports by using machine learning algorithms. \citet{saire2020text} considered a case study to analyze the publications in Mexico and examined people's behavior. Most available work on COVID-19 data, however, presented on ``exploratory data analysis'' such as standard word clouds and histograms of most posted words. In this article, we conduct sentiment analysis \citep[e.g.,][]{pak2010twitter,agarwal2011sentiment,kouloumpis2011twitter} to understand the impact of COVID-19 on the emotion of the public.

While there has been some discussion about the public emotional reaction to the \mbox{COVID-19} pandemic using the text mining techniques based on the available platforms, the existing research is subject to several limitations: (1) most studies conducted sentiment analysis based on a single word and ignoring the interaction among words \citep[e.g.,][]{xue2020public}.  For example, completely opposite meaning from the original word is expressed if this word is combined with the word ``not" or ``no"; (2) most current studies \citep[e.g.,][]{lwin2020global,pastor2020sentiment} only present the results without discussing its connection to the anti-epidemic measures; (3) emotions of individuals change over time and differ from place to place due to different levels of anti-epidemic measures but such features are not necessarily incorporated in the most available studies; (4) only few studies \citep[e.g.,][]{zhou2020examination} consider the possible typos, slangs, word variation, abbreviation, and emoji, which commonly appear in the casual environment of Twitter.


In this article, we mainly focus on two aspects of public sentiments regarding the \mbox{COVID-19} pandemic. The first questions is: ``how do people react to the spread of COVID-19 over time"? Specifically, we are interested in comparing the change of emotion in different time periods based on tweets, and we also study the association between people's reactions to the COVID-19 and the number of daily reported infected cases. Second, multiple measures have been implemented by the government to mitigate the virus spread, but it is still unclear how serious people take those measures. We are also interested in people's reactions by using the keyword ``lockdown", ``masks", and ``vaccine" in the analysis. To alleviate possible confounding effects associated underlying factors including culture, government practice, and internet access regulations etc., here we restrict our analysis to the four cities of Canada, Toronto, Montreal, Vancouver, and Alberta, which are mostly hit by COVID-19. We analyze sentiments expressed in four cities of Canada on one of the most widely used social media platform, Twitter – a microblog website, for the period of February 24, 2020, to October 14, 2020.

The remainder of the article is organized as follows. In Section 2, we discuss the study design and the sentiment analysis method, which includes the procedure of text mining using the Twitter data. In Section 3, we present the results of sentiment analysis using Twitter text data of Canada. We compare the results over different periods. Further, we extend the comparisons by contrasting the results for the four cities in the United States. In Section~4, statistical models are built to analyze how the public emotion is associated with the daily reported infected cases. Finally, we conclude the article in Section~5.

\section{Methods} \label{sec:Method}

\subsection{Study Design}

\begin{figure}[h]
 \centering
 \includegraphics[width=6in]{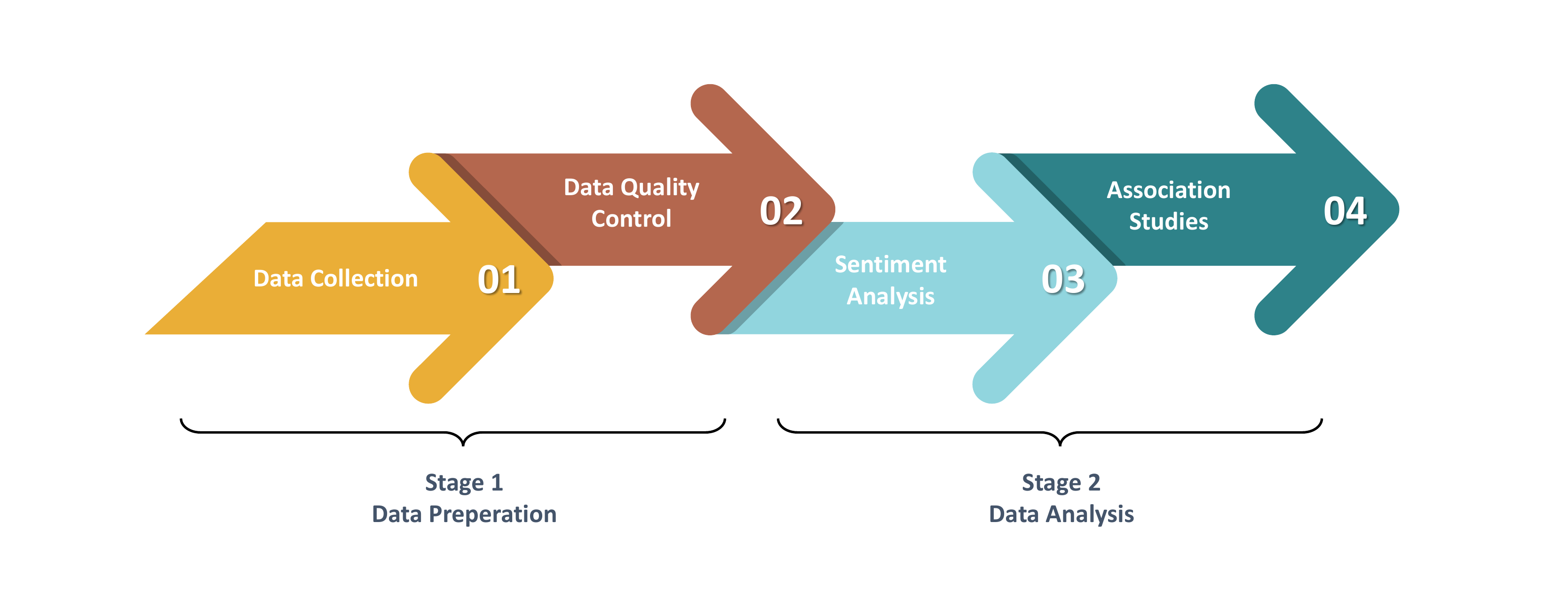}
 \caption{A flowchart of Twitter data mining pipeline}\label{fig:flowchart}
\end{figure}

Our Twitter data mining pipeline consists of a data preparation stage and a data analysis stage, shown in Figure~\ref{fig:flowchart}.  The data preparation procedure includes data collection and raw data cleaning. \mbox{COVID-19} related tweets, published between February 24, 2020 and October 14, 2020, are collected from four Canadian cities: Toronto, Montreal, Vancouver, and Calgary, the representative cities of the four most populated provinces, Ontario, Quebec, British Columbia, and Alberta, respectively. Four cities in the U.S.: New York, Los Angeles, Seattle, and Chicago, are considered for comparison.   We use the {\bf snscrape} module on {\sf Python~3.8} (\url{https://github.com/JustAnotherArchivist/snscrape}) to scrape the tweet text online in the search of using the keyword ``COVID-19".  All the appearing tweets that are published within 50$km$ of the center of each considered city are included in our analysis, leading to 30,655 tweets in total for the four cities in Canada and 69,742 tweets in total for the four cities in the United States. The data cleaning procedure is to be described in detail in the following section; and the clean data are then analyzed using {\bf pandas} package of {\sf Python~3.8}. 

The data analysis stage includes the construction of descriptive statistics characterizing  the topic popularity and public sentiment. We specifically compare the sentiment scores within different regions dynamically with temporal effects incorporated. Furthermore, we build a statistical model to study the association among the sentiment scores, Twitter activities, and the number of reported infected cases.

\subsection{Data Cleaning}

Applying the common standards \citep[e.g.,][]{stone2011comparing,xue2020public}, we first clean raw data using the {\bf pandas} module of {\sf Python~3.8}, following \citet{xue2020public}:
\begin{itemize}
    \item[(1)] The URLs, hashtag symbol ``$\#$" and symbol ``@" are removed from the tweets in the data set;
    \item[(2)] The tweets written in a non-English language are removed;
    \item[(3)] Meaningless characters, punctuation, and stop-words are removed from the dataset as they do not contribute semantic meanings.  Some examples are shown in Table~\ref{tab:meaningless}.
\end{itemize}

\begin{table}[htbp]
\centering
\caption{An example list of meaningless characters, punctuation, and stop-words \label{tab:meaningless}}
\begin{tabular}{rrr}
\hline
\textbf{Character} & \textbf{Punctuation} & \textbf{Stop-words} \\ \hline
**                  & ,                    & I                   \\
SSS                  & ()                    & me                  \\
=                  & ;                    & myself              \\
\$                 & -                    & during              \\
\&                 & $\sim$               & before              \\ \hline
\end{tabular}
\end{table}

\subsection{Sentiment Analysis}\label{sec:sentanalysis}

Tweets are primarily connected with moods and sentiments. To reflect this feature, we carry out sentiment analysis, which is a process of identifying an attitude of the author on a topic that is being written about.

The evaluation of the sentimental effect is conducted by matching the words to the existing lexicons \citep{Jagdale2018} and look up for their sentiment score in the lexicon. We firstly break each tweet into individual words, determine the sentiment score of each word in the tweet, and then obtain the sentiment score for each tweet by adding those scores for each word. Different lexicons have been proposed in the field of  text mining. Some of the lexicons are purely polarity-based \citep[e.g.,][]{liu2012survey}, i.e. the sentiment is classified as \textit{negative}, \textit{neutral}, or \textit{positive}. Some lexicons, e.g., AFINE proposed by \citet{AFINN2011} and the Valence Aware Dictionary and sEntiment Reasoner (Vader) proposed by \citet{gilbert2014vader}, take the intensity of the sentiment into account, where the sentiment intensity score is a scale ranged between a negative and a positive values. NRC Sentiment and Emotion Lexicons \citep{mohammad2013nrc}  consider each word to be one or a combination of multiple moods, including \textit{anticipation}, \textit{positive}, \textit{negative}, \textit{sadness}, \textit{disgust}, \textit{joy}, \textit{anger}, \textit{surprise}, \textit{fear}, and \textit{trust}. Table~\ref{Tab:laxicon} shows examples of the lexicons of different types.

\begin{table}[htbp]
\centering
\caption{Examples of different lexicons: Vader, Bing and NRC}
\label{Tab:laxicon}
\begin{tabular}{lr|ll|ll}
\hline
\multicolumn{2}{c}{Vader} & \multicolumn{2}{c}{Bing} & \multicolumn{2}{c}{NRC} \\ \hline
word           & score     & word         & sentiment & word        & sentiment  \\ \hline
:-)       & 1.3        & abominable   & negative  & abacus      & trust      \\
lmao 	      & 2.0        & abominably   & negative  & abandon     & fear       \\
lol       & 2.9        & abominate    & negative  & abandon     & negative   \\
abducted       & -2.3        & abomination  & negative  & abandon     & sadness    \\
abduction      & -2.8        & abort        & negative  & abandoned   & anger      \\
agrees         & 1.5         & admonition   & negative  & accident    & negative   \\
alarm          & -1.4        & adorable     & positive  & accident    & sadness    \\
alarmed        & -1.4        & adore        & positive  & accident    & surprise   \\
alarmist       & -1.1        & adored       & positive  & accidental  & fear       \\
amaze	      & 2.5        & adorer       & positive  & accidental  & negative   \\
amort	           & -2.1        & adoring      & positive  & accidental  & surprise   \\ \hline
\end{tabular}
\end{table}

In this article, we use Vader and NRC to characterize the sentiment and emotion of the tweets, respectively. The Vader lexicon quantifies the sentiment into numeric scores which can be used for further analysis, and the NRC lexicon provides detailed categories to describe the mood in a refined fashion. The Vader lexicon is also specifically attuned to sentiments expressed in social media. It contains the utf-8 encoded emojis and emoticons, which are important features frequently used in the tweet texts \citep{gilbert2014vader}. To conduct sentiment analysis, polarity scores are identified for every single word in each tweet according to the Vader lexicon, and the frequency of each mood in emotion categories appearing in the text are identified according to the NRC lexicon.  For the sentiment score obtained by Vader, we calculate the average sentiment score of each tweet by first summing the scores of each words in the tweet and then dividing by the total number of words in the tweet. Following \citet{gilbert2014vader},  after calculating the sentiment scores, we further adjust the calculated tweet-wise sentiment scores to incorporate the information related to negation words (``not" and ``n't"), punctuation to intensify sentiments  (e.g., ``Good!!!"), conventional use of word-shape to signal emphasis (e.g., using CAPITAL words), the word modifying the intensity (e.g., ``very", ``pretty", etc.), and the conventional slangs and emoji (e.g, ``lol",``:)"). For the case with ``??", we treat it as an intensified punctuation just like ``!!", but for the case with ``?", we do not made any adjustment because of the uncertainty of being a real question or a rhetorical question. After computing the positive, neutral and negative scores for each tweet, we further calculate the compound score according to the rules in \citet{gilbert2014vader}, and then normalize it to be between -1 (most extreme negative) and +1 (most extreme positive). These steps are implemented via the \texttt{polarity\_scores()} function in the Vader module (\url{https://github.com/cjhutto/vaderSentiment}).

For the mood categories annotated by the NRC lexicon, the accumulating number of counts of each emotion category for each tweet is also calculated. We further compute the frequency, defined as the ratio of the counts of each emotion to the total word count in a tweet using the \texttt{NRClex()} function in the \textsf{python} NRCLex module (\url{https://github.com/metalcorebear/NRCLex}). For each word, the emotion is represented by a ten-dimensional vector to reflect 10 different moods specified in Section~\ref{sec:sentanalysis}, where each element is expressed as the frequency of a mood, ranging from 0 (extremely lack of this emotion) to 1 (extremely full of this emotion). As an example, sentiment analysis of the tweets on February 24, 2020 in Toronto, Canada is presented in Tables~\ref{tab:exampletweets1} and \ref{tab:exampletweets2}.

\begin{table}[htbp]
\tiny
\centering
\caption{Examples of COVID-19 related tweets collected in Toronto, on February 24, 2020\label{tab:exampletweets1}}
\begin{tabular}{ll}
\hline
No.                & Tweet                                                                                                        \\ \hline
\multirow{2}{*}{1} & The trajectory of   the coronavirus is unknown at this time and it's possible that cases have                \\
                   & \multicolumn{1}{c}{occurred in other countries that don't have the proper tools to diagnose and contain it.} \\
2                  & GIS real-time data map of coronovirus   COVID19 from john hopkins                                            \\
3                  & Fantastic thread by our Dr. Hota   describing how the system works to protect Canadians                      \\
4                  & ...and flights are still letting people in?                                                                  \\
5                  & passenger on montreal   to vancouver aircanada flight tested positive for coronavirus COVID-19               \\ \hline
\end{tabular}
\end{table}

\begin{table}[htbp]
\setlength\tabcolsep{1.5pt}
\tiny
\centering
\caption{Sentiment scores of the example tweets\label{tab:exampletweets2} in Table~\ref{tab:exampletweets1}}
\begin{tabular}{lrrrrrrrrrrrrrrr}
\hline
                   & \multicolumn{4}{c}{Vader Lexicon}                                                                                        & \multicolumn{1}{c}{} & \multicolumn{10}{c}{NRC Lexicon}                                                                                                                                                                                                                                                                       \\
No.                & \multicolumn{1}{c}{Negative} & \multicolumn{1}{c}{Neutral} & \multicolumn{1}{c}{Positive} & \multicolumn{1}{c}{Compound} & \multicolumn{1}{c}{} & \multicolumn{1}{c}{Anticipation} & \multicolumn{1}{c}{Positive} & \multicolumn{1}{c}{Negative} & \multicolumn{1}{c}{Sadness} & \multicolumn{1}{c}{Disgust} & \multicolumn{1}{c}{Joy} & \multicolumn{1}{c}{Anger} & \multicolumn{1}{c}{Surprise} & \multicolumn{1}{c}{Fear} & \multicolumn{1}{c}{Trust} \\ \hline
\multirow{2}{*}{1} & \multirow{2}{*}{0}           & \multirow{2}{*}{1}           & \multirow{2}{*}{0}          & \multirow{2}{*}{0}           &                      & \multirow{2}{*}{9}               & \multirow{2}{*}{8}           & \multirow{2}{*}{8}           & \multirow{2}{*}{8}          & \multirow{2}{*}{8}          & \multirow{2}{*}{7}      & \multirow{2}{*}{8}        & \multirow{2}{*}{8}           & \multirow{2}{*}{8}       & \multirow{2}{*}{7}        \\
                   &                              &                              &                             &                              &                      &                                  &                              &                              &                             &                             &                         &                           &                              &                          &                           \\
2                  & 0                            & 1                            & 0                           & 0                            &                      & 2                                & 2                            & 3                            & 1                           & 3                           & 1                       & 2                         & 1                            & 2                        & 2                         \\
3                  & 0                            & 0.721                        & 0.279                       & 0.7351                       &                      & 3                                & 4                            & 5                            & 4                           & 2                           & 3                       & 3                         & 2                            & 3                        & 4                         \\
4                  & 0                            & 1                            & 0                           & 0                            &                      & 3                                & 4                            & 4                            & 4                           & 4                           & 4                       & 4                         & 3                            & 4                        & 3                         \\
5                  & 0                            & 0.769                        & 0.231                       & 0.5574                       &                      & 2                                & 2                            & 2                            & 2                           & 2                           & 2                       & 2                         & 2                            & 2                        & 3                         \\ \hline
\end{tabular}
\end{table}


\section{Descriptive Statistics}\label{sec:des-stat}

Before conducting text mining of tweets, we summarize demographic information of the tweets data scraped online. The number of tweets reflects the popularity of a topic on Twitter, and ``like",  ``replies" and  ``retweets" are three main activities for users to engage with the tweet, whose counts are an indication of the impact of a tweet in generating discussions. In Figures~\ref{fig:demo1} and \ref{fig:demo2}, we present the city-level trajectory of the number of tweets and of the total number of \textit{likes}, \textit{replies}, and \textit{retweets} for the COVID-19 associated tweets, together with that of the daily number of reported infected cases for the province or states in which the city is located. 

Figures~\ref{fig:demo1} and \ref{fig:demo2} show that the numbers of like, of reply and of retweet have a similar trend, which are fairly similar to the number of tweets in the overall pattern. Those trajectories exhibit a reasonable commonality within a country, and the peaks for all the cities in Canada and the U.S. appear nearly at the same time - around the middle of March, 2020. Interestingly, those peaks come prior to the peaks of the number of reported infected cases, which may reflect that in the early stage of the virus spread in North America, the \mbox{COVID-19} outbreak in other countries in Asia and Europe has already stimulated a large number of active discussions. 

Figure~\ref{fig:sentimentVader} presents a heatmap of the daily average sentiment score for the eight cities over time, where the darkness of the color shows the magnitude of the sentiments, with orange color and green color representing positive and negative sentiments, respectively. The heatmap is constructed using the \texttt{geom\_tile()} function in the \textsf{R} package ggplot2. The sentiment change varies from city to city, not just from country to country. For the period between the end of February and early March, except for Los Angeles, all the sentiments in all other cities show overall negative patterns. Between the middle of March and early June, all the cities show positive sentiments though the degree varies. After June, sentiment reaction fluctuate noticeably from city to city, with all cities having varying negative patterns in September, which is likely related to the second wave of a surging trend of COVID-19 cases. Overall, Toronto and New York remain consistently positive between the middle of March and the beginning of September, whereas the other cities exhibit varying negative sentiments intermittently.

To trace the change of sentiment scores in different cities over time, we summarize the mean and standard deviation of tweetwise sentiment scores, stratified by city and the time periods as in Figure~\ref{fig:timeline}. Table~\ref{tab:ttwise-ss-period} reports on the results for the cities in Canada and U.S.A. All the four Canadian cities and four U.S. cities have the largest sentiment scores for Period~2 (i.e., during the lockdown), which may indicate confidence and positive feelings about COVID-19 during the lockdown period. In contrast, smaller mean scores in Period~1 than in Period~2 may be related to the concern and the uncertainty about the disease in the early stage of the pandemic. The sentiment scores in U.S. is more negative than those in Canada in all the periods. While the mean scores differ in cities, the associated standard deviations remain similar for different cities and different periods.

To closely understand how anti-epidemic measures may be related to the daily average sentiment scores (obtained from Vader approach) of the tweets,  we produce the heatmaps for three keywords: ``mask", ``vaccine", and ``lockdown", and display them in Figures~\ref{fig:sentimentVader-mask}--\ref{fig:sentimentVader-lockdown}, respectively. Opinions on ``mask" are generally positive in Toronto but changes from time to time in other three Canadian cities. Except for Seattle, opinions on ``mask" appear to be positive or neutral in other three U.S. cities. On the other hand, people generally tend to have negative opinions on ``vaccine" in most cities. For ``lockdown", opinions are generally positive from March to May when most cities are in the lockdown status or emergency states (Figure~\ref{fig:timeline}), whereas most Canadian cities (Vancouver, Montreal, and Calgary) and two US cities (Seattle and Chicago) have negative sentiments after reopening. Among the three anti-epidemic measures, ``vaccine" has the most long-lasting negative views, showing the lack of confidence in the development of the vaccine.

To closely visualize the change of the mood composition of the daily tweets over time, in Figure~\ref{fig:sentimentNRC}, we present the density plots obtained from the NRC method. Overall, the changes in Canadian cities seem to be more variable than U.S. cities, and the trend and trajectory vary from city to city and from time to time. For example, from May to September, the moods are slightly intensified in Vancouver, Montreal, and Calgary, whereas an opposite trend is observed for Toronto and New York. Relative to other months, February is the month that incur a large variation of the word frequency in each mood for most cities, which may be attributed to the uncertainty and the lack of knowledge of COVID-19 in the early stage of the pandemic.

\section{Causal Inference} \label{sec:causal}

\subsection{Examination Framework} \label{sub:exam-frame}
With the descriptive analysis reported in Section~\ref{sec:des-stat}, we are further interested in examining the Vader sentiment scores in terms of the potential causal relationships.  Specifically, here we explore possible causal relationships of the daily number of reported infected cases with tweet activities (e.g., tweet total counts, tweet like counts, tweet reply counts, and retweet counts) and with sentiment scores of COVID-19 related tweets. We implement convergent cross mapping \citep{ye2015distinguishing}, integrated with the echo state network approach \citep{lukovsevivcius2009reservoir}, to explore the causal relationships.

Let $X$ and $Y$ denote the two variables whose causal relationship is of interest. It is our goal to identify whether there is evidence to suggest a causal relationship between $X$ and $Y$ by examining $\{X_t: t=1,\ldots,T\}$ and $\{Y_t: t=1,\ldots,T\}$, the time series of observations for $X$ and $Y$, respectively, where $X_t$ and $Y_t$ are observed values of $X$ and $Y$ at time $t$, respectively.  For example, $X_t$ represents the daily average sentiment score on day $t$ and $Y_t$ can be the number of tweets on day $t$.

The idea of the convergent cross mapping is that if $Y$ is the cause of $X$, then the time series $\{Y_t:t=1,\ldots,T\}$ of the causal variable $Y$ can be recovered from the time series $\{X_t:t=1,\ldots,T\}$ of the effect variable $X$ \citep{Tsonis2018}. To facilitate this rationale and possible the lag effects, let $\tau$ denote the lag time, and we take $X_{t}$ as the input data and repeatedly fit the leaky echo state network model \citep{JAEGER2007335} to predict $Y_{t+\tau}$ by varying the value of $\tau$, denoted $\hat{Y}_{t+\tau}$. The detail is given in Section~\ref{sec:Echo-State-Network}. 

Now we describe the evaluation of the performance of the prediction of $Y_{t+\tau}$ using the $X_{t}$.  For any given $\tau\le T$, we compute the Pearson's correlation coefficient between the predicted time series $\widehat{Y}_{t+\tau}$ and their corresponding observations $Y_{t+\tau}$ \citep{ye2015distinguishing}, denoted $\rho_y(\tau)$ and given by
\begin{equation*}
    \rho_{y}(\tau) = \frac{\sum_{t=1+|\tau|-h(\tau)}^{ T-h(\tau)}Y_{t+\tau}\widehat{Y}_{t+\tau}-\sum_{t=1+|\tau|-h(\tau)}^{ T-h(\tau)}Y_{t+\tau}\sum_{t=1+|\tau|-h(\tau)}^{ T-h(\tau)}\widehat{Y}_{t+\tau}}{\sqrt{\sum_{t=1+|\tau|-h(\tau)}^{ T-h(\tau)}(Y_{t+\tau}-\bar{Y}_{t+\tau})^2\sum_{t=1+|\tau|-h(\tau)}^{ T-h(\tau)}(\widehat{Y}_{t+\tau}-\bar{\widehat{Y}}_{t+\tau})^2}},
\end{equation*}
where $\bar{Y}_{t+\tau} = \frac{1}{T-|\tau|}\sum_{t=1+|\tau|-h(\tau)}^{ T-h(\tau)}Y_{t+\tau}$, $\bar{\widehat{Y}}_{t+\tau} = \frac{1}{T-|\tau|}\sum_{t=1+|\tau|-h(\tau)}^{ T-h(\tau)}\widehat{Y}_{t+\tau}$, and 
\begin{equation*}
    h(\tau)=\begin{cases}
        \tau,  & \text{if } \tau \ge 0, \\
        0, & \text{if } \tau < 0.\\
    \end{cases}
\end{equation*}

Likewise, taking $Y_{t}$ as the input data and $X_{t+\tau}$ as output labels, we fit another echo state network model in predicting the values of $X_{t+\tau}$, denoted as $\widehat{X}_{t+\tau}$. {Then, for any given $\tau$,} we compute the Pearson's correlation coefficient between $\widehat{X}_{t+\tau}$ and $X_{t+\tau}$, denoted as $\rho_x(\tau)$.

Following the lines of \citet{ye2015distinguishing} and \citet{huang2020detecting}, we determine the causality between X and Y according to the time lag $\tau$ so that $\rho_{x}$ and $\rho_{y}$ reach their peak values. Specifically,
\begin{itemize}
    \item If $X$ causes $Y$ and not vice versa, we expect the peak value of the $\rho_{x}(\tau)$ to be located in the positive domain, i.e., the corresponding $\tau$ is positive. Meanwhile, we expect the peak value of the $\rho_{y}(\tau)$ to be at a negative $\tau$.
    \item  If $X$ and $Y$ cause each other, both the peak values of the $\rho_{x}(\tau)$ and the $\rho_{y}(\tau)$ are expected occur in the negative domain.
    \item If the coupling of $X$ and $Y$ has a delay effect, the lag positions of the peaks of the $\rho_{x}(\tau)$ and the $\rho_{y}(\tau)$ are expected to be influenced by the delay time.
\end{itemize}

To study how $\rho_{x}(\tau)$ and $\rho_{y}(\tau)$ may change at different time lag $\tau$, we consider the range $[-30,30]$ for $\tau$ in the analysis in Section~\ref{sec:COVID-example}.

\subsection{Echo State Network}\label{sec:Echo-State-Network}

As discussed in Section~\ref{sub:exam-frame}, the identification of the causality relationship between $X$ and $Y$ is carried out by examining the Pearson correlation coefficient between the observed data, say $Y_{t+\tau}$, and their predicted counterparts, say $\widehat{Y}_{t+\tau}$, which are obtained by fitting a prediction model connecting $X$ and $Y$. While various modeling schemes may be considered for building a prediction model, here we employ the echo state network approach for its good performance in the prediction of non-linear time series data \citep{lukovsevivcius2009reservoir}. 

\begin{figure}[h]
 \centering
 \includegraphics[width=5in]{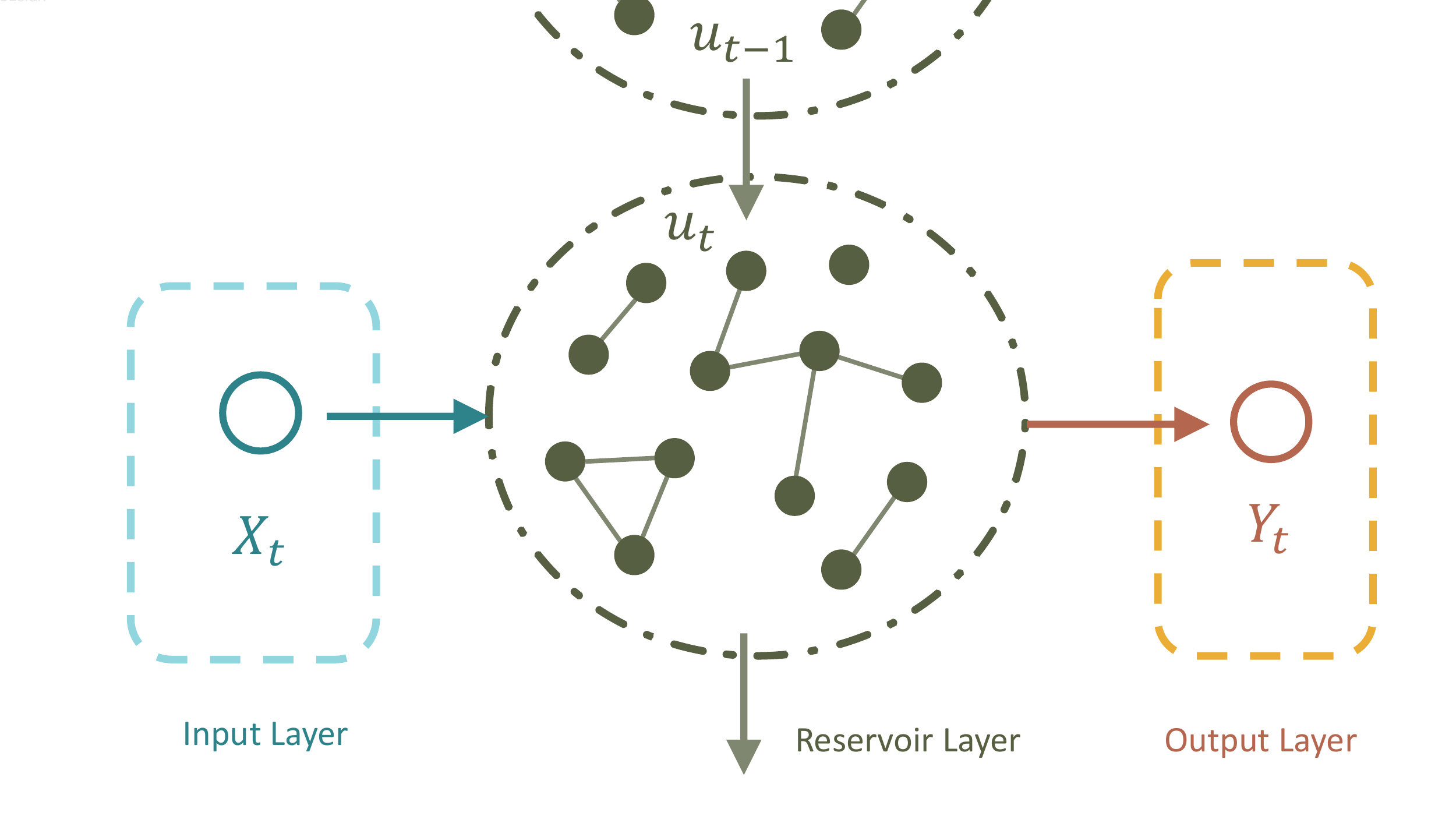}
 \caption{The diagram of a leaky echo state network}\label{fig:echo-state-network}
\end{figure}

The echo state network is basically composed of three layers: the input layer, the reservoir layer, and the output layer (Figure~\ref{fig:echo-state-network}). The input layer contains the input data and the output layer is made up of the output labels. The reservoir layer consists of the hidden reservoir neuron states, denoted $u_t$, which form an $N\times1$ vector with $N$ being a user-specified positive integer. An $N\times N$ adjacent matrix, say $A$, is constructed to describe the connections among the reservoir neurons. 

The determination of the links among the three types of layers is carried out with two procedures: the input-to-reservoir procedure and the reservoir-to-output procedure, which are described as follows using the data:

\begin{itemize}
    \item From the input layer to the reservoir layer: 
    
    For any $t=1,\ldots, T$, we construct the relationships between the impute data $X_t$ and the layers of hidden reservoir neuron states. Let $u_{t}^{(0)}$ denote the initial reservoir neuron states, then we update the neuron states $u_t$ by 
    \begin{align*}
        u_t^* &= \text{tanh} \{ A u_t^{(0)} + W_{\rm in} X_{t}\}; \\
        u_t &= (1-\psi) u_{t-1} + \psi u_t^*,
    \end{align*}    
    where $\psi$ is the leaky parameter to be specified, and $W_{\rm in}$ is an $N\times 1$ matrix representing the weight to transform $X_t$ from input layer to reservoir layer. Here, $\text{tanh}(v)=\frac{\exp(v)-\exp(-v)}{\exp(v)+\exp(-v)}$.
    
    \item From reservoir layer to output layer: 
    
    We compute the predicted output labels,
    \begin{equation*}
        \widehat{Y}_{t+\tau} = W_{\rm out} u_t,
    \end{equation*}
    where $W_{\rm out}$ is an $n\times N$ matrix representing the weight matrix transform $u_t$ into output layer.
\end{itemize}

In these steps, only the weights $W_{\rm out}$ need to be trained by minimizing the penalized loss function
\begin{equation*}
   \norm{ \widehat{Y}_{t+\tau} - Y_{t+\tau}}_2+ \alpha\norm{ W_{\rm out}}_2,
\end{equation*}
where $\norm{\cdot}_2$ represents the $L_2$-norm and $\alpha$ is the tuning parameter to be specified.

The adjacent matrix $A$ and the weight matrix $W_{\rm in}$ are prespecified, which may be specified using the following procedure. Let $p_s$ be a user-specified value between 0 and 1, let $\gamma$ be a user-determined positive scaling parameter determining the degree of nonlinearity in the reservoir dynamics \citep{goudarzi2015exploring}; and let $\beta$ be a pre-specified positive scaling parameter. Generate a sequence of values, say $v$, independently from Uniform $[-1,1]$ and a sequence of values, say $s$, independently from Bernoulli($p_s$), and then we form the weight matrix $W_{\rm in}$ by letting its elements be given by $\gamma sv$. The adjacent matrix $A$ is formed similarly with its element given by $\beta sv$. To ensure the echo state network to work properly, the effects of initial conditions should vanish as the times series evolves \citep{jarvis2010extending}, which is also known as the echo state property. A necessary condition to ensure the echo state property is that the largest eigenvalue of $A$, denoted $\lambda_{\rm max}$, is smaller than 1 \citep{JAEGER2007335}. Here, $\lambda_{\rm max}$, also called the  ``spectral radius" of $A$, determines the time range that the time series data interact each other non-linearly. Such a property guides us to set a suitable value for scale parameter $\beta$.


Treating variable $X_t$ as the training data and $Y_{t+\tau}$ as the output labels, we implement the echo state network using the {\sf EchoTorch} module \citep{echotorch} in {\sf Python 3.8}. To determine suitable values of the tuning parameters $\lambda_{\rm max}$, $\psi$, $N$, $p_s$, $\alpha$, and $\gamma$, we take the leave-one-out cross-validation procedure. Specifically, we take the time series of seven cities as the training data and the remaining one city as the testing data. The procedure is repeated by leaving each city out as testing data once, where in each study, we record the normalized root-mean-squared-error (NRMSE)  \citep{lukovsevivcius2009reservoir}:
\begin{equation*}
    \text{NRMSE} =  \frac{\sqrt{\sum_{t=1}^T(\widehat{Y}_t-Y_t)^2/T}}{\sum_{t=1}^T Y_t/T}.
\end{equation*}
We conduct a grid search to find the optimal set of tuning parameters such that the NRMSE are minimized.

Similarly, to study another direction of the causal relationship, the procedure described above is repeated by switching $X$ and $Y$. 

\subsection{Analysis Results}\label{sec:COVID-example}

In this study, we examine possible causal relationships among the six features pairwisely: COVID-19 daily infected cases; Vader sentiment scores; the daily number of tweets, and the daily total number of likes, of retweets and of replies. Assuming that the causal relationships among those features are the same for each city, we merge the data in all cities and consider two directions of the relationship as explained in Section~\ref{sub:exam-frame}. 

For illustrations of the implementation of the procedures described in Sections~\ref{sub:exam-frame}--\ref{sec:Echo-State-Network}, in Direction 1, let $\{X_t:t=1,\ldots, T\}$ denote the time series of the daily average sentiment scores in a city  and let $\{Y_t:t=1,\ldots, T\}$ denote the time series of the daily tweet counts of that city; and in Direction 2, we swap the $X_t$ and $Y_t$. According to the leave-one-out cross-validation, we choose the tuning parameter $\lambda_{\rm max}$, $\psi$, $N$, $p_s$, $\alpha$, and $\gamma$, respectively, to be 0.1, 0.5, 150, 0.1, 0.1, 0.9 in Direction 1, and 0.1, 0.9, 250, 0.7, 100 and 0.9 in Direction 2. Then using those identified optimal parameter values, we train the echo state network for $W_{\rm out}$ to the time series $X_t$ and $Y_t$ to predict $\widehat{Y}_t$ and $\widehat{X}_t$ for both directions, respectively. The predictions are repeated for all the cities considered in the study. Finally, for each $\tau \le T$, we compute the Pearson's correlation coefficient between $\widehat{Y}_{t+\tau}$ and $Y_{t+\tau}$, where $\widehat{Y}_{t+\tau}$ is the predicted value corresponding to $Y_{t+\tau}$. In another direction, we compute the correlation coefficient between $\widehat{X}_{t+\tau}$ and $X_{t+\tau}$. We present the analysis results in Figure~\ref{fig:tauplot}, which shows that in Direction 1, the Pearson's correlation coefficient $\rho_x$ reaches the peak at $\tau=8$, whereas in Direction 2, the Pearson's correlation coefficient  $\rho_y$ is peaked at $\tau=-5$. This suggests that the sentiment of COVID-19 is likely to cause the changes in the time series of the tweet counts, but not vice versa. On the other hand, the small value of the peak of the correlations (i.e., $\rho_x=0.006$ and $\rho_y=0.169$), indicate a weak relationship between the infected case number and the tweet counts.

We repeat the analysis for each pair of features of our interest and identify the optimal lag that achieves the highest correlation in each scenario, shown in Figure~\ref{fig:tauplotsall}, which reveals three findings: (i) the infected cases and COVID-19 sentiment scores have an instantaneous bidirectional relationship as $\tau=0$ for both direction; (ii) the sentiment scores unidirectionaly cause the changes of the tweet-related features, including tweet, like, reply and retweet counts; (iii) the tweet activity features, including tweet, like, reply, and retweet, are instantaneously related to each other. The messages in (ii) and (iii) are not surprising, but the message (i) appears somewhat counter-intuitive. It is expected that the change of the number of infected cases can influence people's sentiment regarding COVID-19, but not necessarily vice versa. However, closely examining the  magnitude from the direction of sentiment scores to infected cases, it is not as strong as the other way around. People's sentiments about COVID-19 may influence their personal activities which may, in turn, contribute more risk of infection. The results of the causal relationships are summarized in Figure~\ref{fig:causalchart}.

\section{Discussion} \label{sec:discussion}

The ongoing COVID-19 pandemic has presented tremendous challenges to the public and it has significantly affected our daily life. It is important to study public sentiments in reacting to anti-epidemic measures and evaluate the impact of the pandemic on the public mental health. In this paper, we conduct sentiment analysis of the tweets in eight North American cities  on the topics of the COVID-19 spread, masks, lockdown, and vaccine. We apply the echo state network and convergent cross-mapping to characterize possible causal relationships among the relevant features. 

Our visualization of sentiment scores provides intuitive information about people's reaction to the virus spread as well as the implemented anti-epidemic measures.  The descriptive statistics reveal that the public sentiments vary in different periods of the pandemic. For example, the sentiments become positive in Period 2 (during lockdown) after being negative in Period 1 (before lockdown) due to the uncertainty and lack of knowledge about the disease. The sentiments also vary in different cities, not just in different countries. While people are generally thinking positively about the mask usage, the sentimental reactions to the vaccine and lockdown are much negative. Our analysis of the causal relationships shows that the infected cases and COVID-19 sentiment scores are instantaneously correlated. The sentiment influences the activities of Twitter users, such as publish, like, reply, or retweet a tweet, which are also correlated with each other. 

On the other hand, the analysis has the limitation that the full meaning of the sentences is not deciphered, and thereby, care needs to be paid to interpret the negative sentiment results. For instance, when interpreting negative sentiments of the ``mask" related tweets, one needs to consider two possible reasons; the negative opinions may arise from the usage of masks as an anti-epidemic measure or the thoughts on other people who do not follow the rules of mask-wearing. Although the two situations are reflecting completely opposite views of the measures, their sentiment direction is the same and this difference cannot be directly distinguished in the sentiment analysis. A more refined analysis is warranted in future work.

\section*{Funding}

This research was supported by the Natural Sciences and Engineering Research Council of Canada (NSERC). Yi is Canada Research Chair in Data Science (Tier 1). Her research was undertaken, in part, thanks to funding from the Canada Research Chairs Program.\\

\section*{Acknowledgements}

The authors would like to thank reviewers for their useful comments.

\section*{Contributions}

The first two authors lead the project with equal contributions including writing the paper; the last two authors participate in the project with equal contributions.\\

\section*{Competing Interests}

The authors declare that they have no competing interests.\\

\section*{Availability of Data and Material}

All codes and data involved in this paper are available on
request from the authors.\\


\bibliographystyle{apa}
\clearpage
\phantomsection  
\renewcommand*{\bibname}{References}

\addcontentsline{toc}{chapter}{\textbf{References}}

\bibliography{SentiTwi}


\appendix
\section{ Appendix}

\begin{figure}[h]
 \centering
 \makebox[10pt]{%
    \includegraphics[width=0.9\paperwidth]{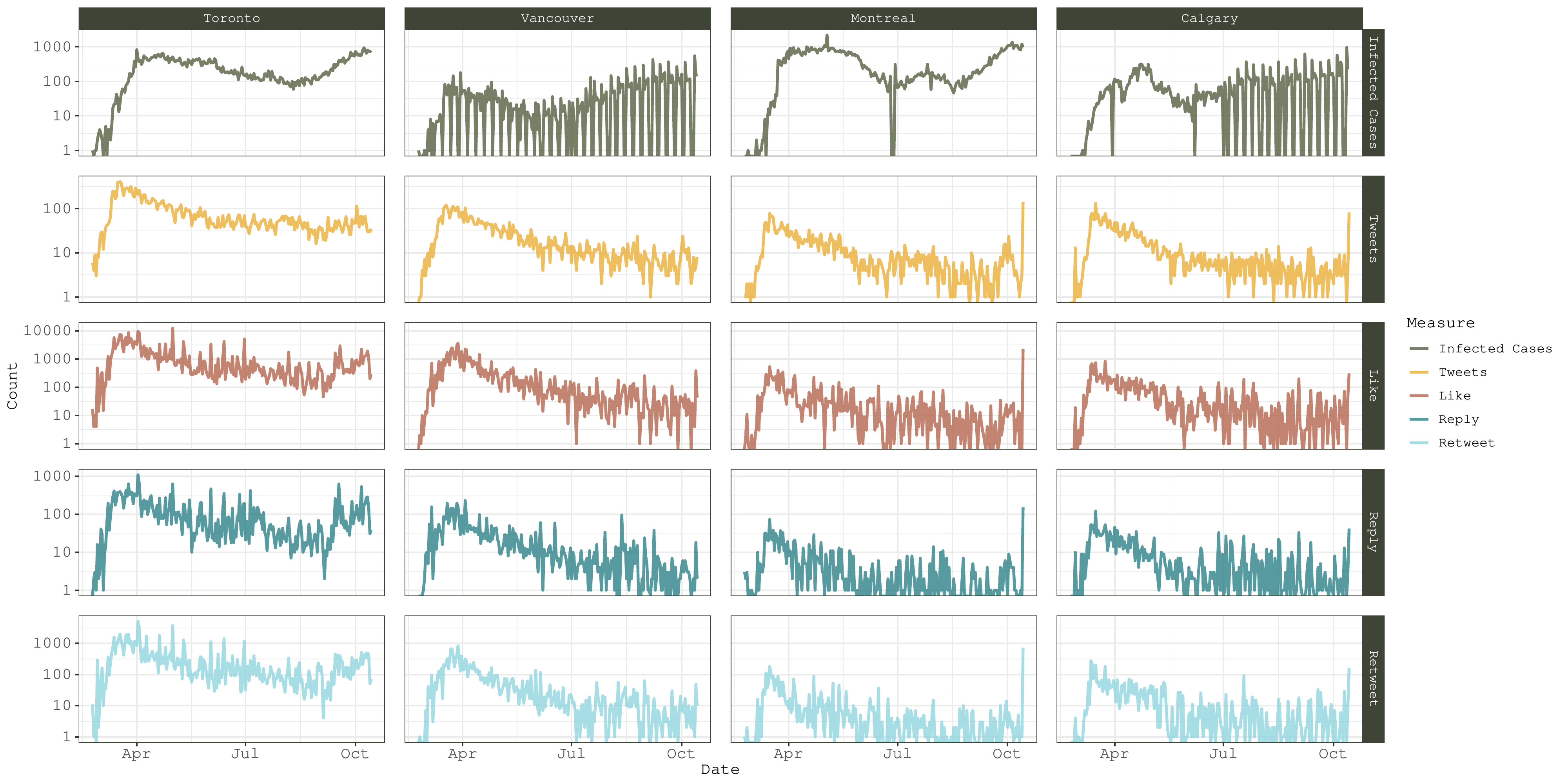}}
 \caption{The trajectory of provincial daily infected cases, daily number of tweets, total counts of likes, reply and retweets of the COVID-19 related tweets for cities in Canada. To properly present the trend, the y-axis is presented with logarithm transformation. }\label{fig:demo1}
\end{figure}

\begin{figure}[h]
 \centering
 \makebox[10pt]{%
    \includegraphics[width=0.9\paperwidth]{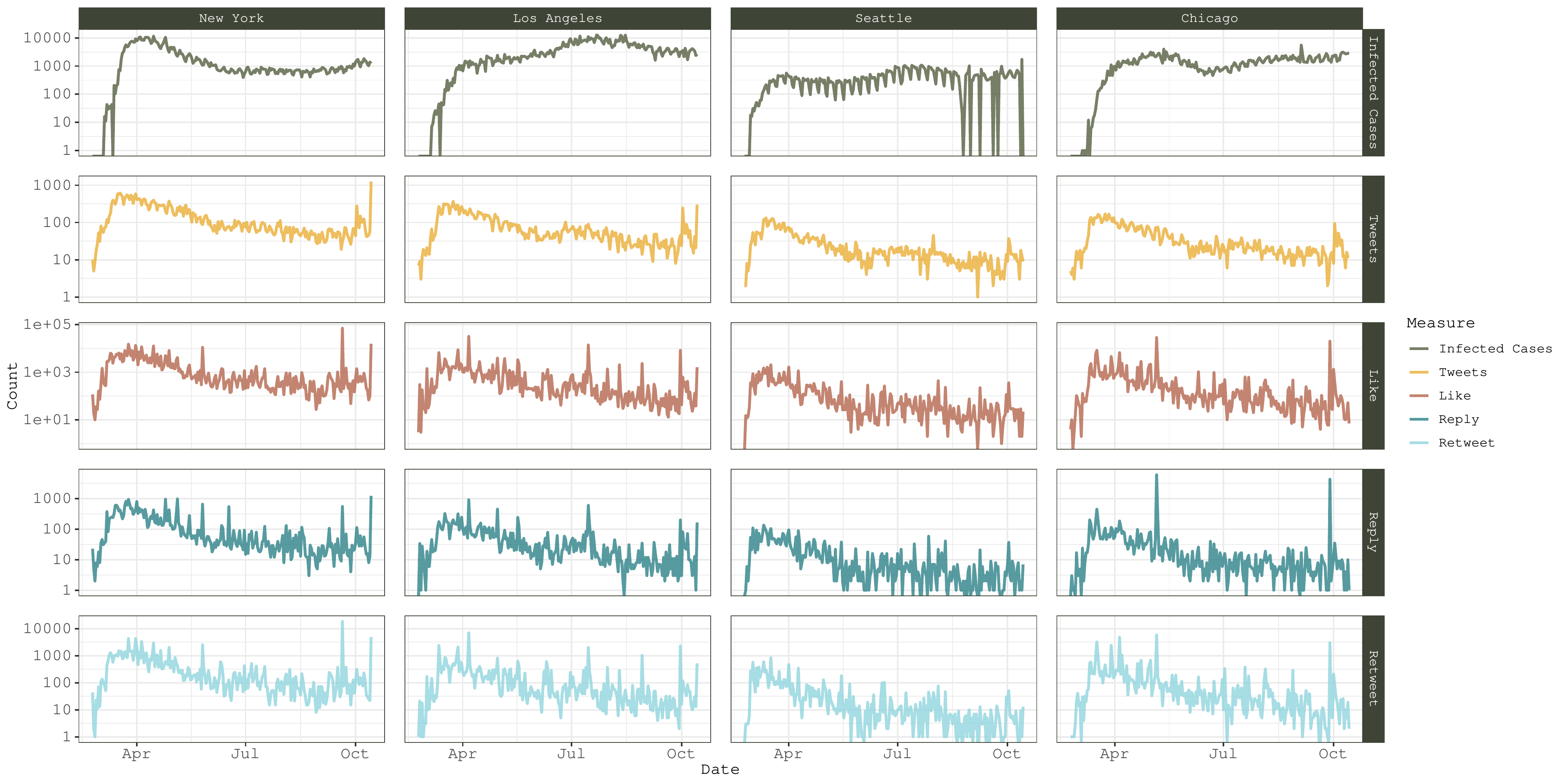}}
 \caption{The trajectory of provincial daily infected cases, daily number of tweets, total counts of likes, reply and retweets of the COVID-19 related tweets for cities in United States. To properly present the trend, the y-axis is presented with logarithm transformation. }\label{fig:demo2}
\end{figure}

\begin{landscape}

\begin{figure}[h]
 \centering
    \includegraphics[width=1.1\paperwidth]{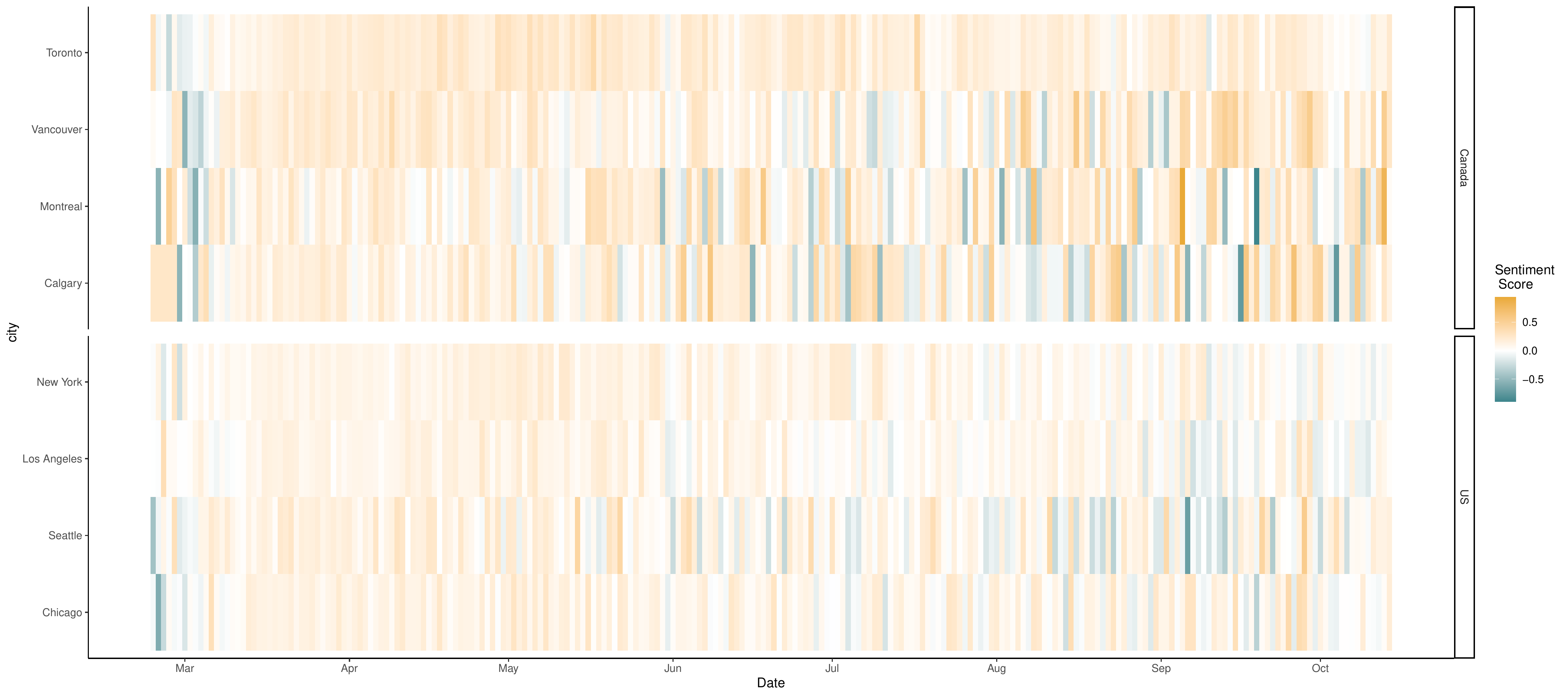}
 \caption{The heatmap of the sentiment score of "COVID19" related tweets calculated using Vader lexicon over time for the eight cities in the North America. The orange color denotes the positive sentiment and the green color represents the negative sentiment.  }\label{fig:sentimentVader}
\end{figure}

\begin{figure}[h]
 \centering
    \includegraphics[width=1.1\paperwidth]{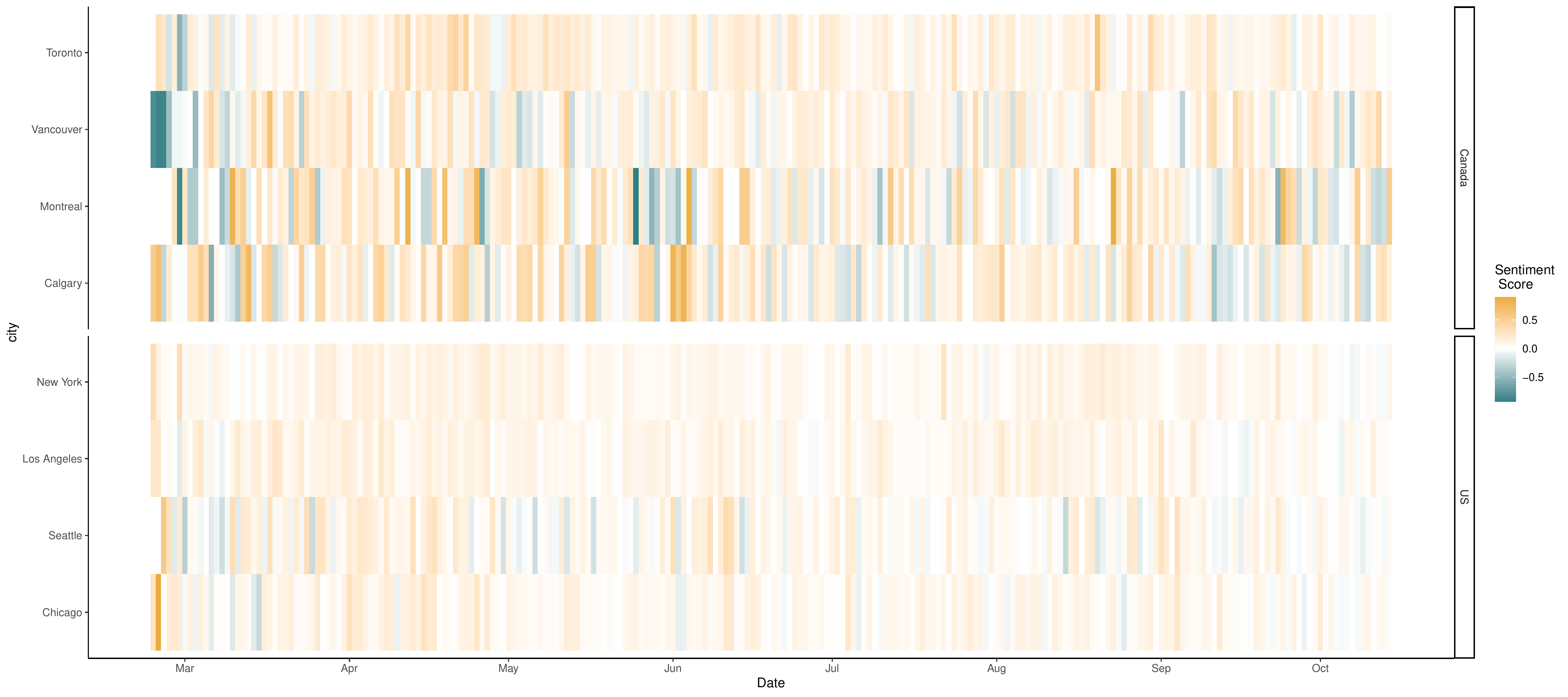}
 \caption{The heatmap of the sentiment score of ``mask" related tweets calculated using Vader lexicon over time for the eight cities in the North America. The orange color denotes the positive sentiment and the green color represents the negative sentiment.  }\label{fig:sentimentVader-mask}
\end{figure}

\begin{figure}[h]
 \centering
    \includegraphics[width=1.1\paperwidth]{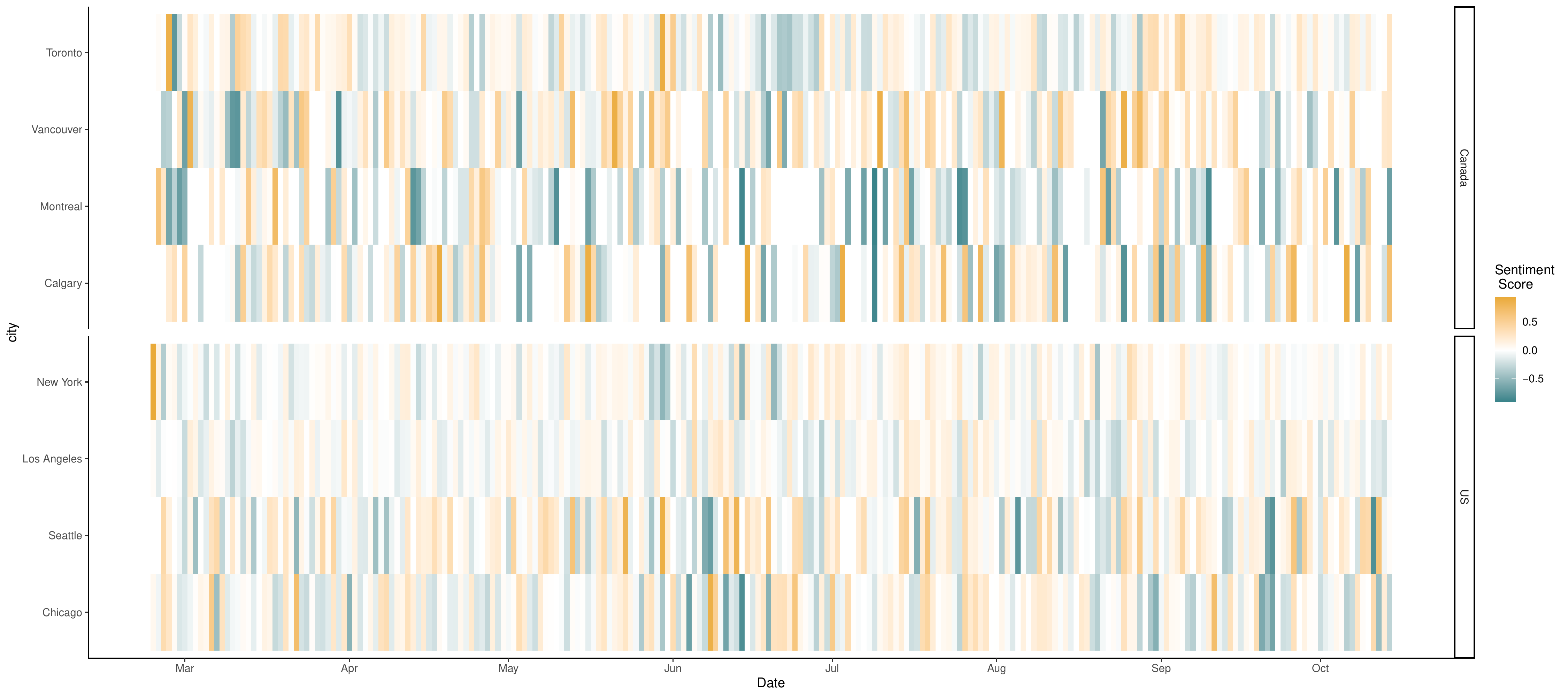}
 \caption{The heatmap of the sentiment score of ``vaccine" related tweets calculated using Vader lexicon over time for the eight cities in the North America. The orange color denotes the positive sentiment and the green color represents the negative sentiment.  }\label{fig:sentimentVader-vaccine}
\end{figure}

\begin{figure}[h]
 \centering
    \includegraphics[width=1.1\paperwidth]{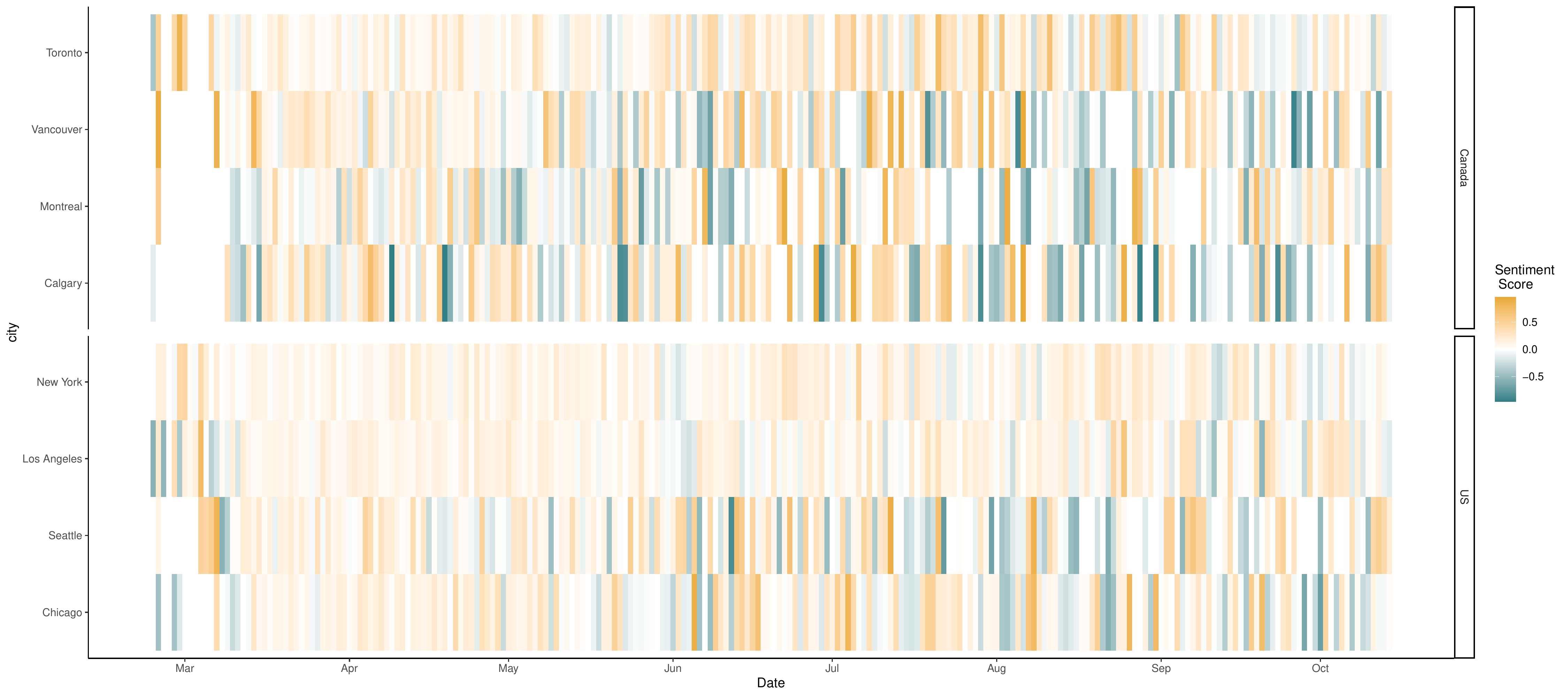}
 \caption{The heatmap of the sentiment score of ``lockdown" related tweets calculated using Vader lexicon over time for the eight cities in the North America. The orange color denotes the positive sentiment and the green color represents the negative sentiment.  }\label{fig:sentimentVader-lockdown}
\end{figure}

\end{landscape}

\begin{table}[htbp]
\setlength\tabcolsep{1.5pt}
\tiny
\centering
\caption{The tweetwise sentiment scores for different cities in different time periods \label{tab:ttwise-ss-period}}
\begin{tabular}{crlrrrrrrrrr}
\hline
\textbf{}                 & \multicolumn{1}{c}{\textbf{}} &  & \multicolumn{4}{c}{\textbf{Canada}}                                                                                      & \multicolumn{1}{c}{\textbf{}} & \multicolumn{4}{c}{\textbf{United States}} \\ \cline{4-7} \cline{9-12} 
Period                    & \multicolumn{1}{c}{Measure}   &  & \multicolumn{1}{c}{Calgary} & \multicolumn{1}{c}{Montreal} & \multicolumn{1}{c}{Toronto} & \multicolumn{1}{c}{Vancouver} & \multicolumn{1}{l}{}          & New York & Los Angeles & Seattle & Chicago \\ \cline{1-2} \cline{4-7} \cline{9-12} 
\multirow{3}{*}{Period 1} & Number of tweets              &  & 581                         & 124                          & 2203                        & 878                           &                               & 614.00   & 129.00      & 50.00   & 196.00  \\
                          & Mean                          &  & 0.12                        & 0.04                         & 0.09                        & 0.13                          &                               & 0.02     & 0.05        & -0.02   & -0.01   \\
                          & s.d.                          &  & 0.46                        & 0.45                         & 0.46                        & 0.49                          &                               & 0.50     & 0.49        & 0.55    & 0.47    \\ \cline{1-2} \cline{4-7} \cline{9-12} 
\multirow{3}{*}{Period 2} & Number of tweets              &  & 1703                        & 1601                         & 9983                        & 3237                          &                               & 25626.00 & 10924.00    & 3655.00 & 5359.00 \\
                          & Mean                          &  & 0.15                        & 0.12                         & 0.20                        & 0.20                          &                               & 0.12     & 0.10        & 0.13    & 0.12    \\
                          & s.d.                          &  & 0.48                        & 0.48                         & 0.49                        & 0.50                          &                               & 0.50     & 0.49        & 0.50    & 0.50    \\ \cline{1-2} \cline{4-7} \cline{9-12} 
\multirow{3}{*}{Period 3} & Number of tweets              &  & 907                         & 1048                         & 7011                        & 1556                          &                               & 9634.00  & 7710.00     & 2188.00 & 3657.00 \\
                          & Mean                          &  & 0.09                        & 0.11                         & 0.15                        & 0.11                          &                               & 0.06     & 0.05        & 0.05    & 0.08    \\
                          & s.d.                          &  & 0.50                        & 0.46                         & 0.47                        & 0.53                          &                               & 0.51     & 0.51        & 0.52    & 0.50    \\ \hline
\end{tabular}
\end{table}

\begin{landscape}

\begin{figure}[h]
 \centering
    \includegraphics[width=1.1\paperwidth]{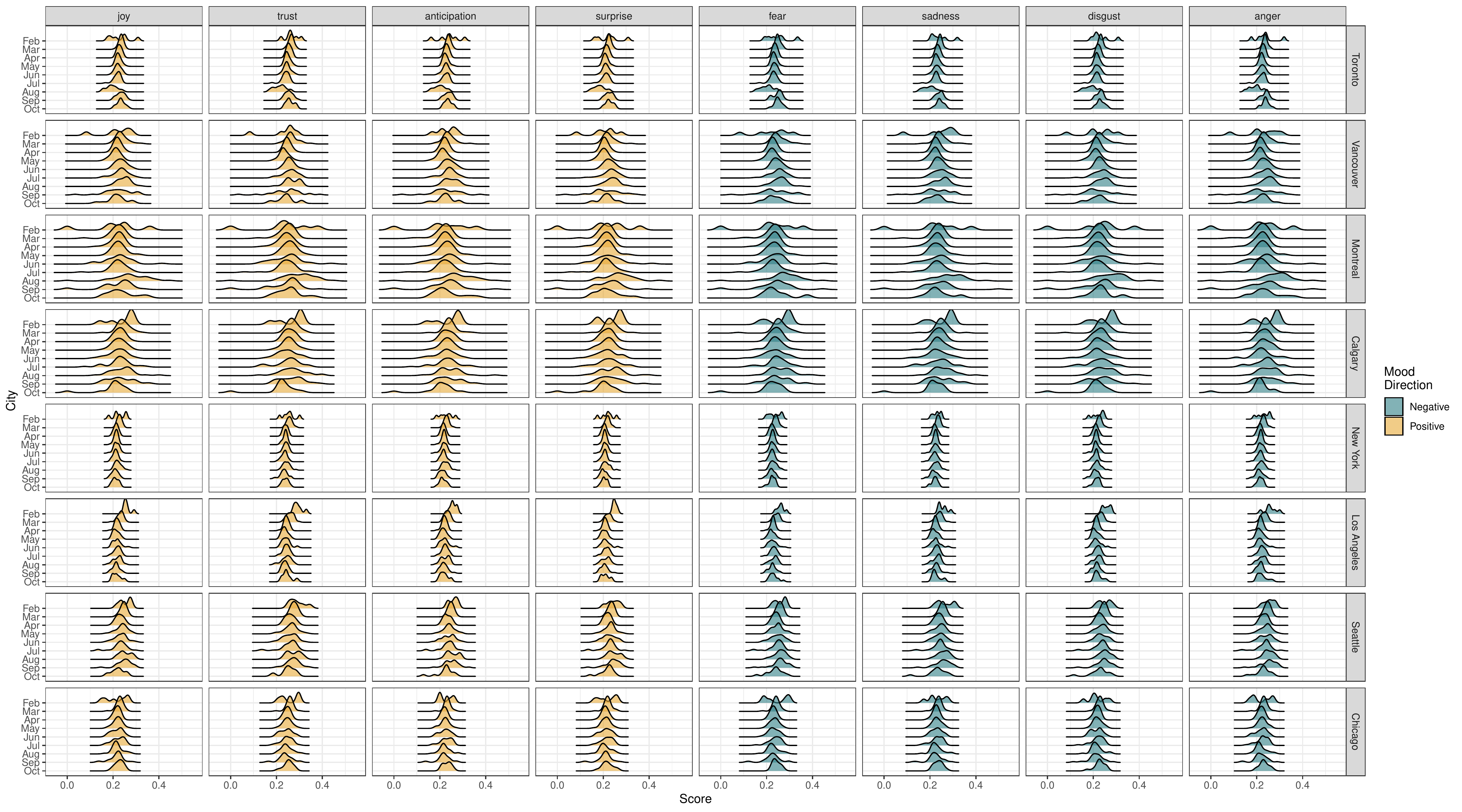}
 \caption{The density graph of the distribution of the moods frequency for different cities over 234 days.  }\label{fig:sentimentNRC}
\end{figure}

\end{landscape}

\begin{figure}[h]
 \centering
 \includegraphics[width=6in]{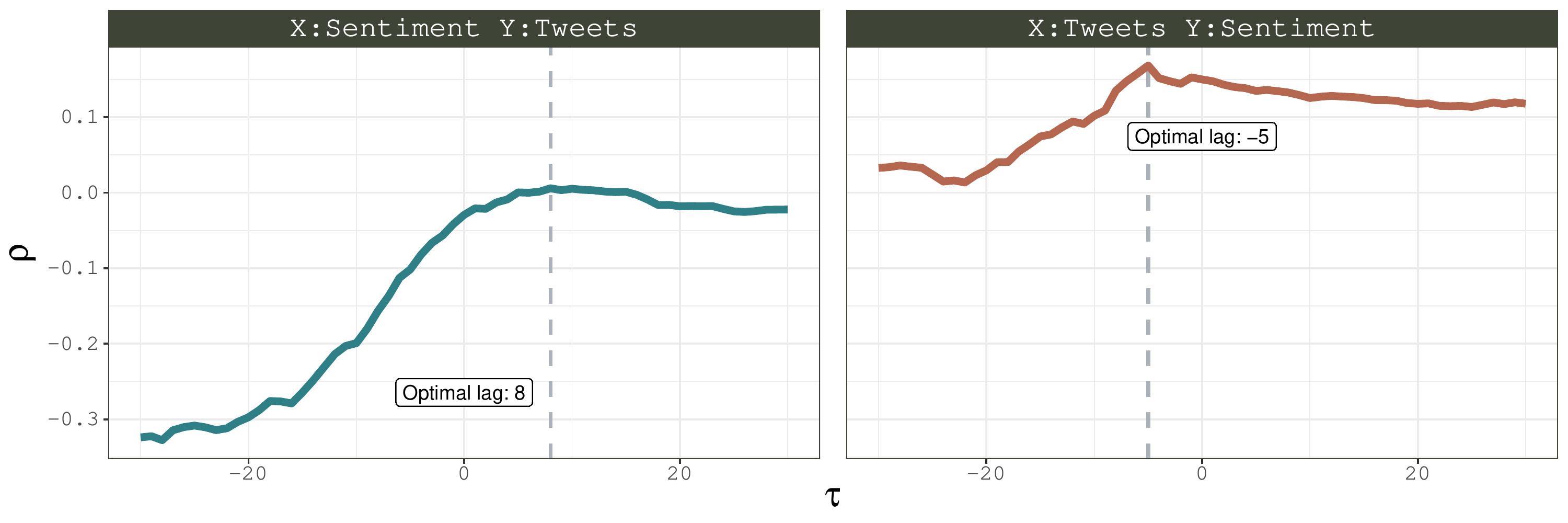}
 \caption{A plot of Pearson's correlation coefficient over different choice of lag in the inference of causal relationship between the infected cases and the tweet counts. The dash vertical line refers to the choice of lag that achieve the peak of the correlation.}\label{fig:tauplot}
\end{figure}

\begin{figure}[h]
 \centering
 \includegraphics[width=6in]{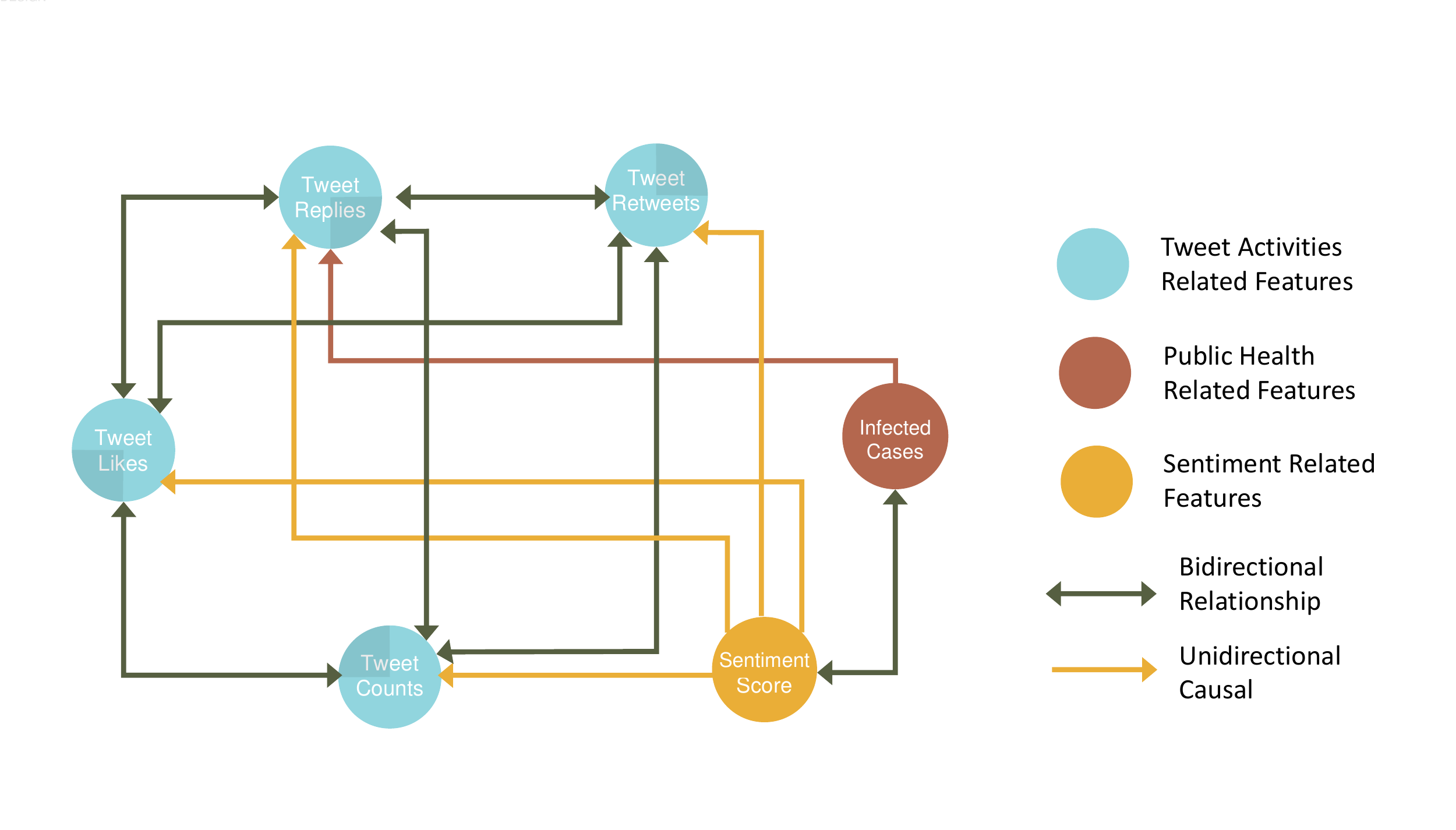}
 \caption{A chart of causal relationships between the tweet activities, infected cases and sentiment scores.}\label{fig:causalchart}
\end{figure}

\begin{figure}[h]
 \centering
 \includegraphics[width=.49\textwidth]{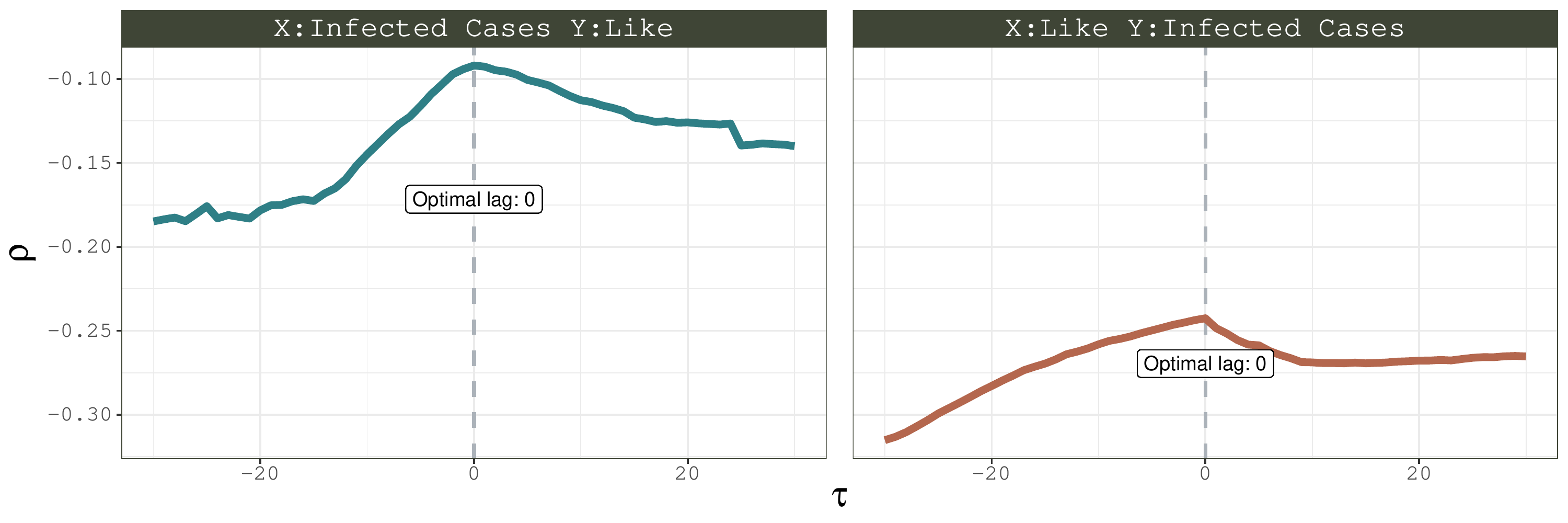}
 \includegraphics[width=.49\textwidth]{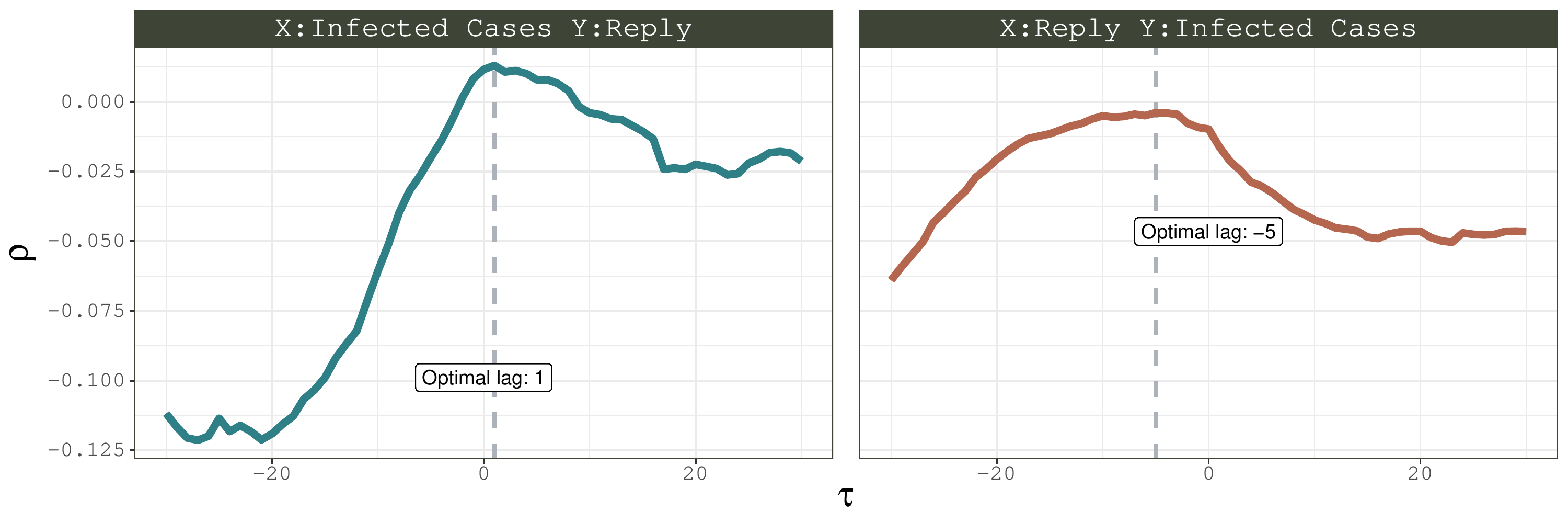}
 \includegraphics[width=.49\textwidth]{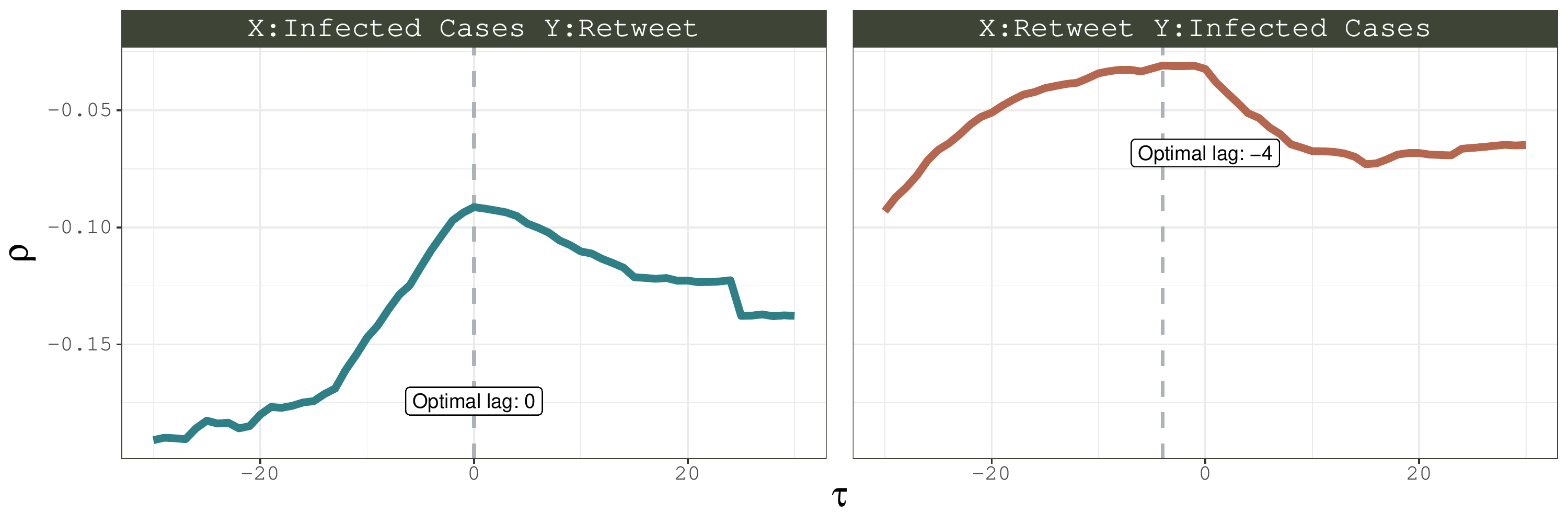}
 \includegraphics[width=.49\textwidth]{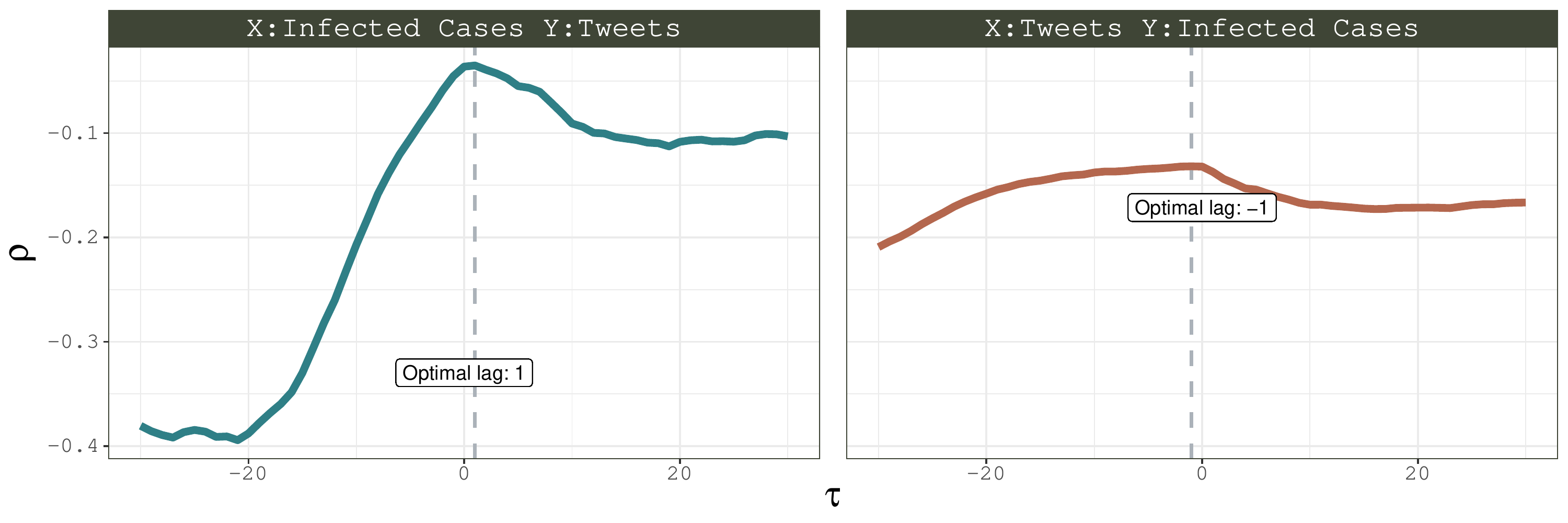}
 \includegraphics[width=.49\textwidth]{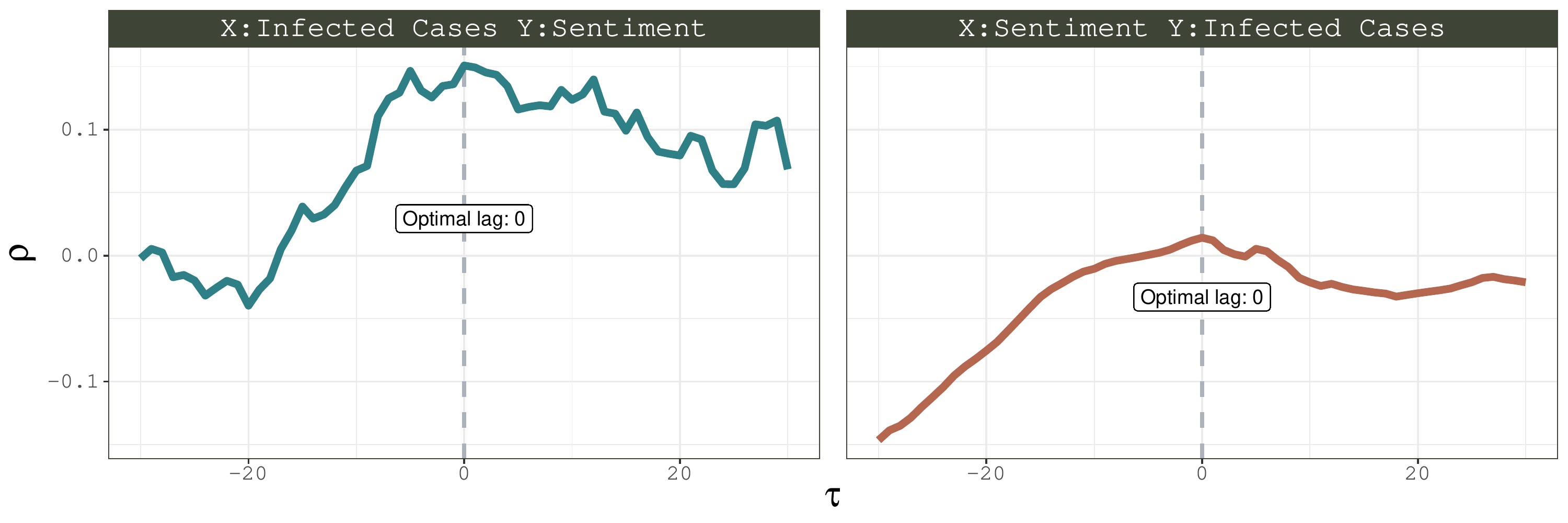}
 \includegraphics[width=.49\textwidth]{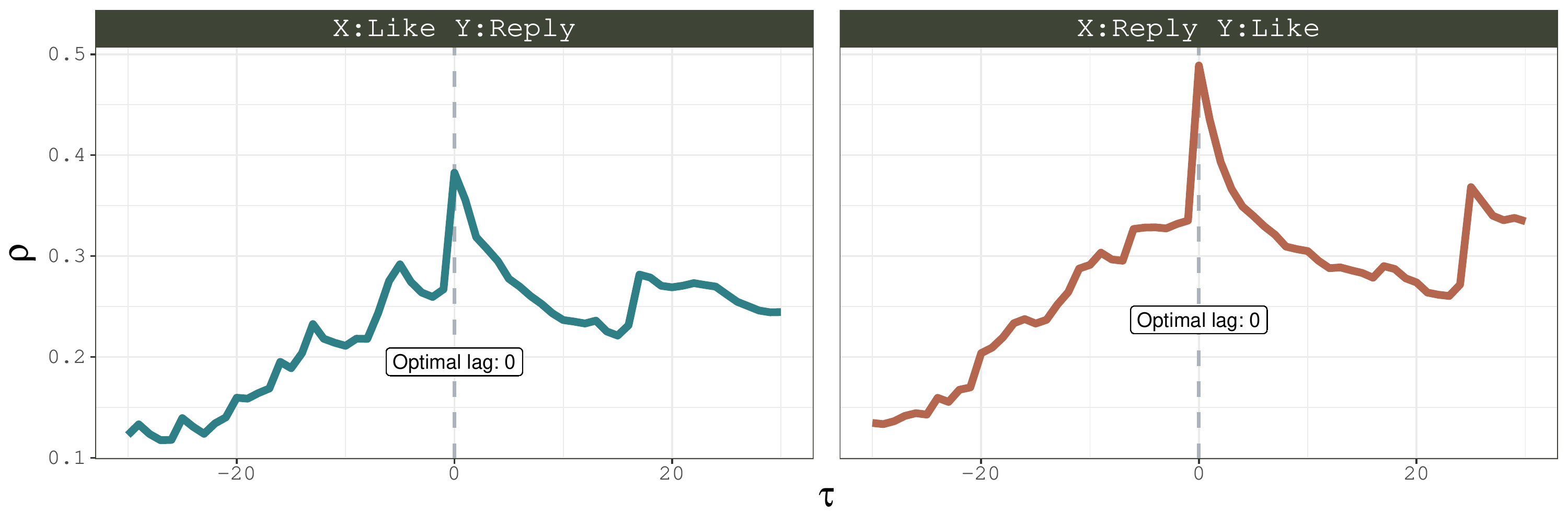}
 \includegraphics[width=.49\textwidth]{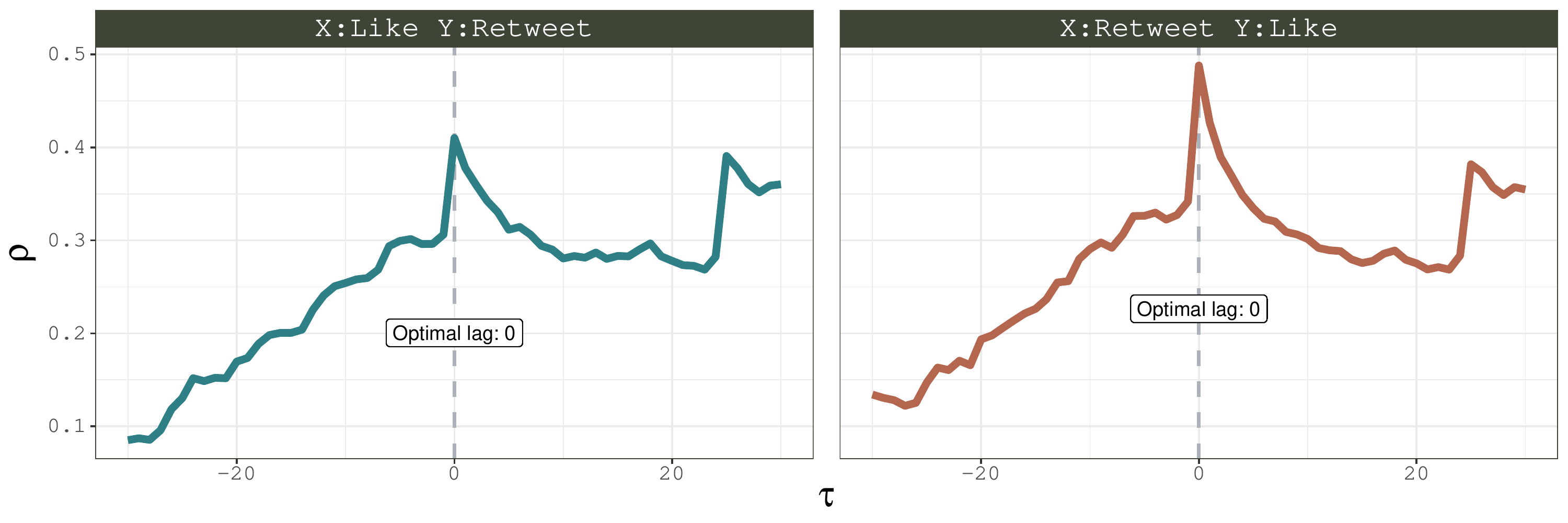}
 \includegraphics[width=.49\textwidth]{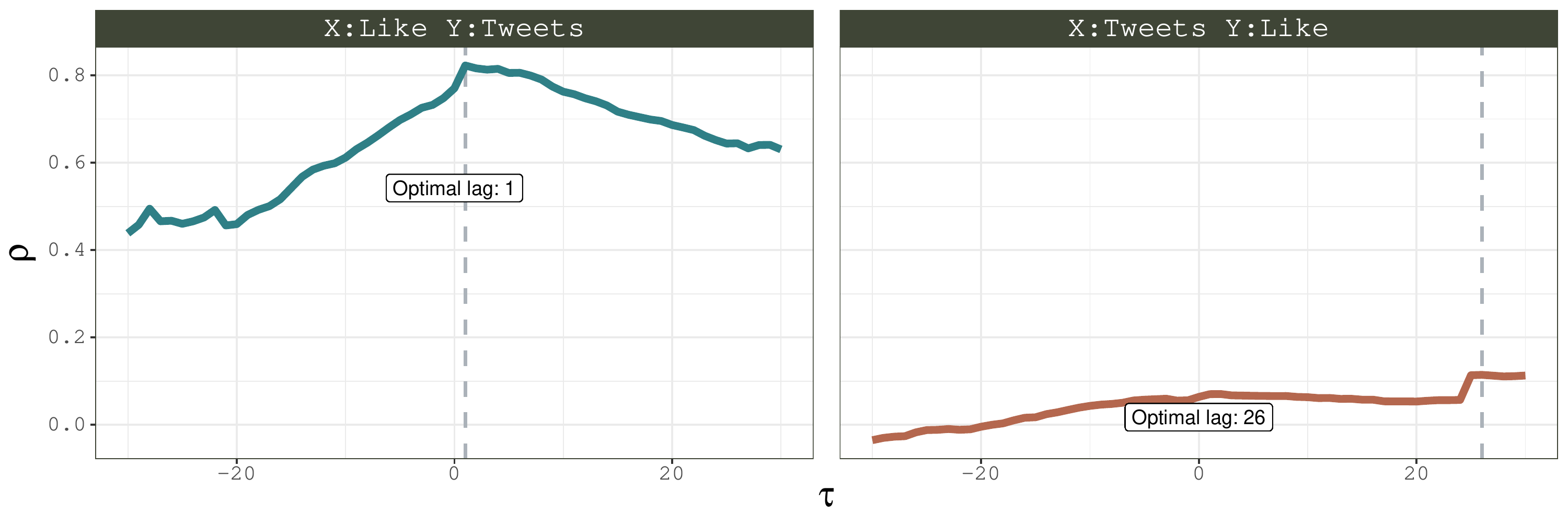}
 \includegraphics[width=.49\textwidth]{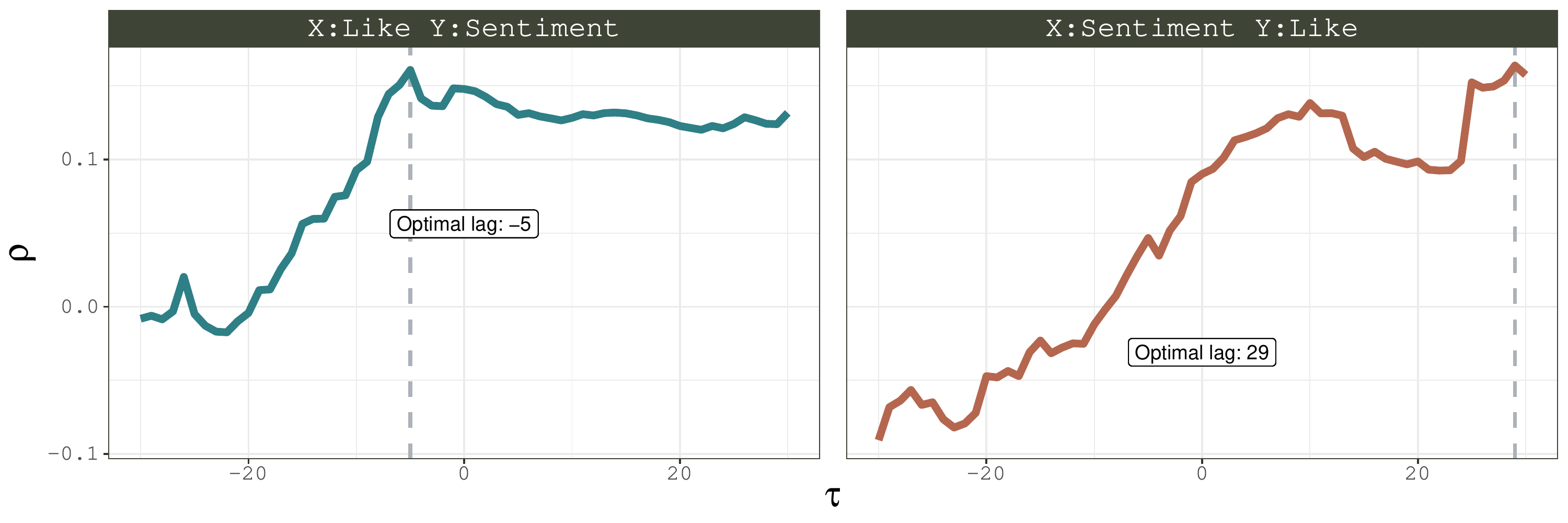}
 \includegraphics[width=.49\textwidth]{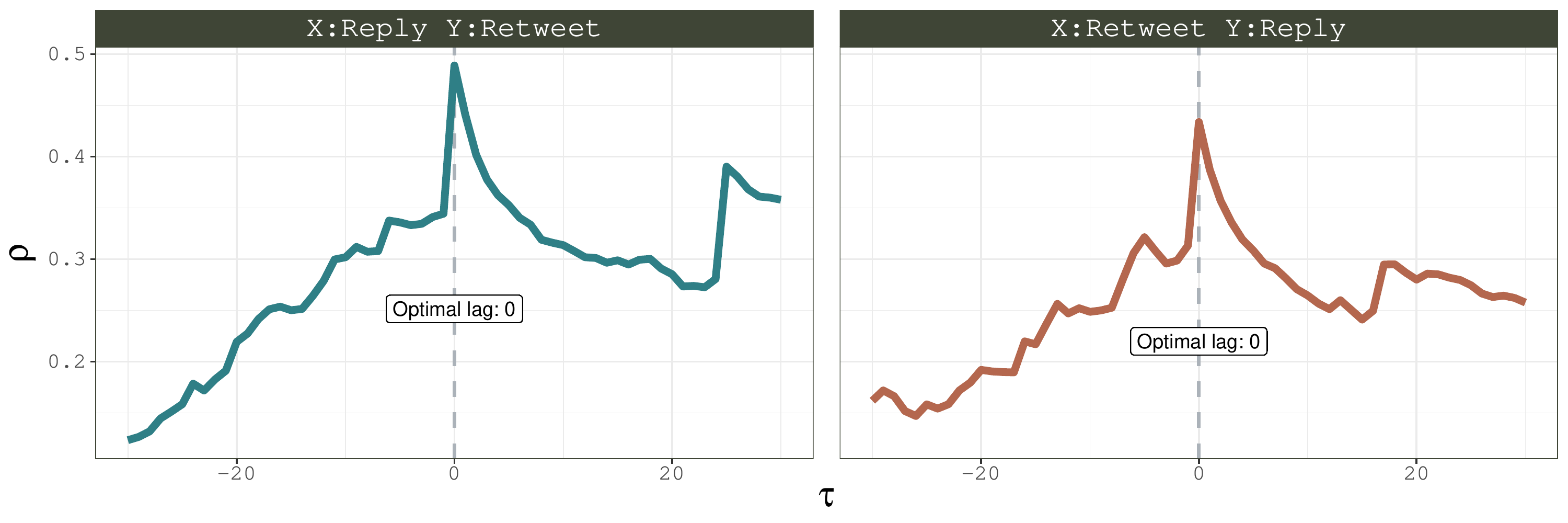}
 \includegraphics[width=.49\textwidth]{figure/tauplot_Like_and_Tweets.pdf}
\includegraphics[width=.49\textwidth]{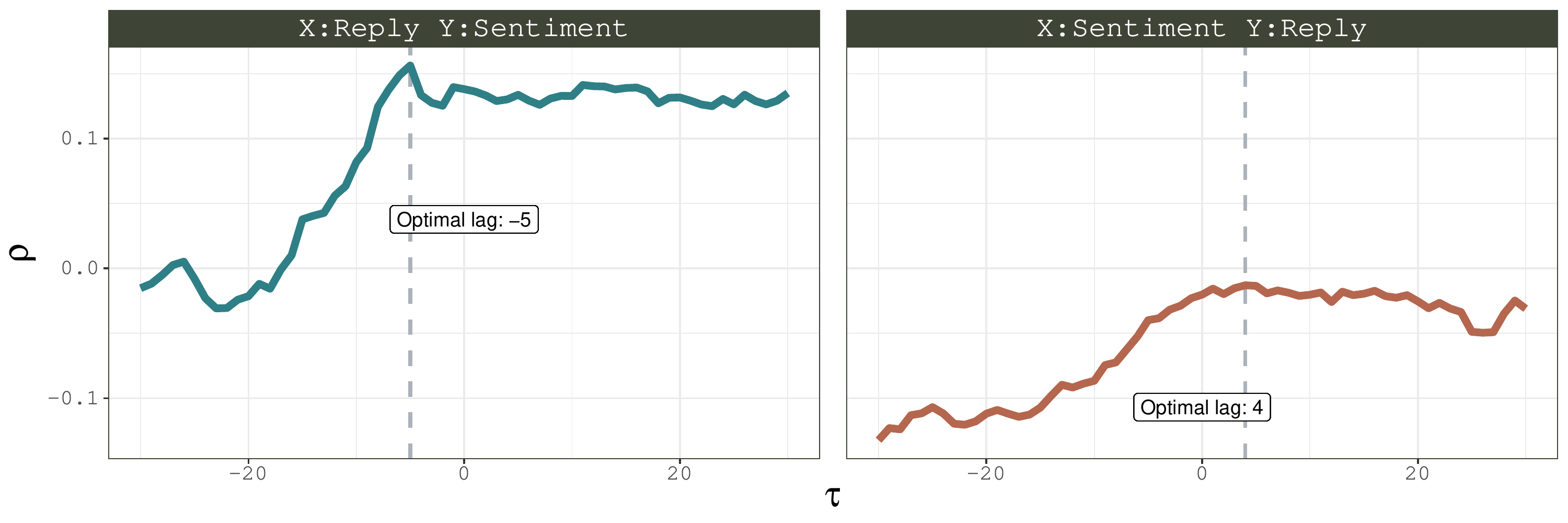}
\includegraphics[width=.49\textwidth]{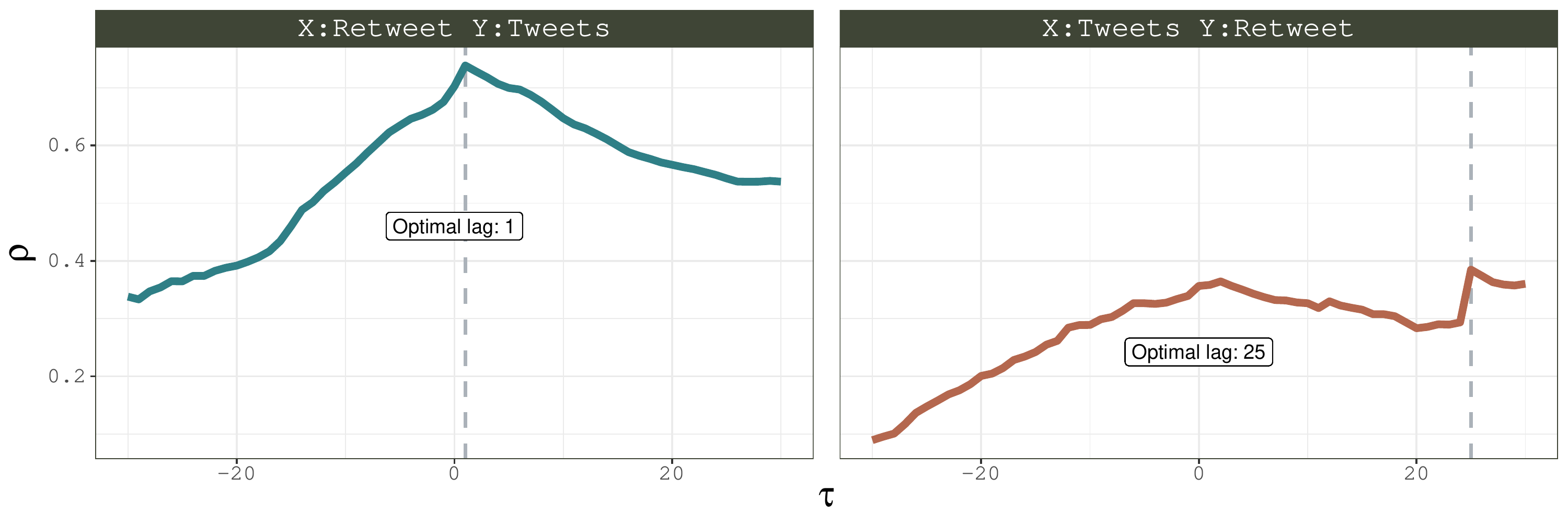}
\includegraphics[width=.49\textwidth]{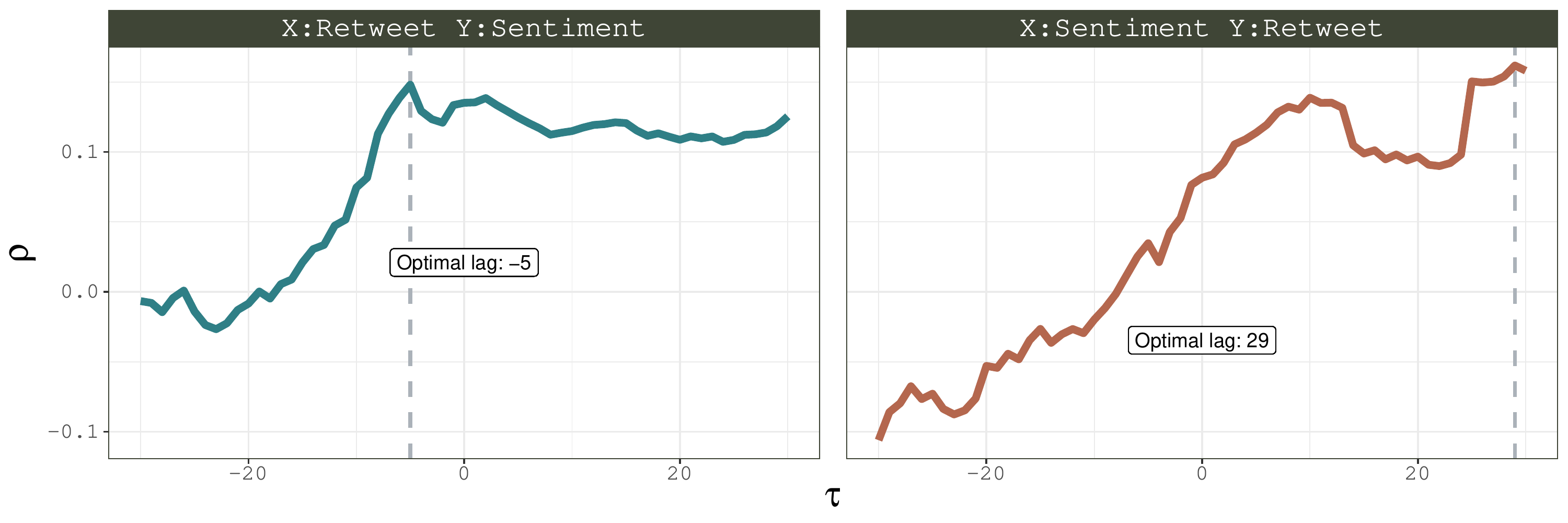}
    
 \caption{Plots of Pearson's correlation coefficient over different choice of lag in the inference of causal relationships. The dash vertical line refers to the choice of lag that achieve the peak of the correlation.}\label{fig:tauplotsall}
\end{figure}

\end{document}